\documentclass[journal,draftcls,onecolumn,12pt]{IEEEtran}
\normalsize
\usepackage{upgreek}
\usepackage{graphicx}
\usepackage{subfigure}
\usepackage{amsmath,bm}
\usepackage{mathrsfs,amsmath}
\usepackage{enumerate}
\usepackage{amssymb}
\usepackage{array}
\usepackage{longtable}
\usepackage{booktabs}
\usepackage{makecell}

\usepackage[mathscr]{euscript}
\usepackage[square, comma, sort&compress, numbers]{natbib}
\usepackage{color}
\usepackage{url}

\usepackage{setspace}

\ifCLASSINFOpdf
\else
\fi

\begin{document}

\title{\huge Reconfigurable Intelligent Surface Assisted Free Space Optical Information and Power Transfer}



\author{Wen~Fang,~Wen~Chen,~\IEEEmembership{Senior~Member,~IEEE,}~Qingqing~Wu,
\IEEEmembership{Senior~Member,~IEEE,}~Kunlun~Wang,~Shunqing~Zhang,~\IEEEmembership{Senior~Member,~IEEE,}
~Qingwen~Liu,~\IEEEmembership{Senior~Member,~IEEE,}~Jun~Li,~\IEEEmembership{Senior~Member,~IEEE}


\thanks{W.~Fang, W.~Chen and Q. Wu are with the Department of Electronic Engineering, Shanghai Jiao Tong University, Shanghai 200240, China (email: wendyfang@sjtu.edu.cn, wenchen@sjtu.edu.cn, qingqingwu@sjtu.edu.cn).}
\thanks{K. Wang is with the School of Communication and Electronic Engineering, East China Normal University, Shanghai 200241, China (email: klwang@cee.ecnu.edu.cn).}
\thanks{S.~Zhang is with Shanghai Institute for Advanced Communication and Data Science, Shanghai University, Shanghai 200444, China (email: shunqing@shu.edu.cn)}
\thanks{Q.~Liu is with the College of Electronic and Information Engineering, Tongji University, Shanghai 200000, China (e-mail: qliu@tongji.edu.cn).}
\thanks{J. Li is with the School of Electronic and Optical Engineering, Nanjing University of Science Technology, Nanjing 210094, China (email: jun.li@njust.edu.cn).}}

\markboth{IEEE Transactions on Communications}%
{Submitted paper}

\maketitle

\vspace{-1.0cm}
\begin{abstract}
Free space optical (FSO) transmission has emerged as a key candidate technology for 6G to expand new spectrum and improve network capacity due to its advantages of large bandwidth, low electromagnetic interference, and high energy efficiency. Resonant beam operating in the infrared band utilizes spatially separated laser cavities to enable safe and mobile high-power energy and high-rate information transmission but is limited by line-of-sight (LOS) channel. In this paper, we propose a reconfigurable intelligent surface (RIS) assisted resonant beam simultaneous wireless information and power transfer (SWIPT) system and establish an optical field propagation model to analyze the channel state information (CSI), in which LOS obstruction can be detected sensitively and non-line-of-sight (NLOS) transmission can be realized by changing the phased of resonant beam in RIS. Numerical results demonstrate that, apart from the transmission distance, the NLOS performance depends on both the horizontal and vertical positions of RIS. The maximum NLOS energy efficiency can achieve $55\%$ within a transfer distance of $10 \rm{m}$, a translation distance of $\pm4 \rm{mm}$, and rotation angle of $\pm50^\circ$.

\end{abstract}
\vspace{-0.5cm}

\IEEEpeerreviewmaketitle

\section{Introduction}\label{Section1}
The rapid commercialization of fifth-generation mobile communication (5G) technology has enhanced communication capacity and expanded not only human connection but also the connection of objects among many terminals~\cite{Shafi2017}. On this basis, the sixth-generation mobile communication (6G) technology has received extensive attention from academia and industry, which presents higher expectations for ultra-wireless broadband, reliable communication, integrated sensing and communication, and energy self-sustainability~\cite{zhang20196g}. However, it is increasingly challenging for the limited wireless spectrum resources to maintain up with the rising demand for wireless spectrum due to the fast expansion of mobile data~\cite{dang2020should}. As a crucial solution for addressing the last mile of wireless access, the free space optical (FSO) system enables the integration of energy transmission, communication, and sensing, utilizing visible light (VL) and infrared bands without requiring additional authorization, with the benefits of no electromagnetic interference and environmentally friendly protection~\cite{zhu2021compensation}. Thus, extensive research has been conducted on simultaneous lightwave information and power transfer (SLIPT) to facilitate high-rate data transfer and uninterrupted power supply for terminals~\cite{uysal2021slipt}.

Among the existing simultaneous lightwave information and power transfer (SLIPT) schemes, the techniques utilizing visible light (VL) and lasers have received extensive research attention~\cite{uysal2021slipt}, \cite{bashir2022energy}. However, the energy received at the receiver using VL is unexpectedly low due to its extremely large spectral bandwidth, which surpasses that of electromagnetic waves by more than 10,000 times~\cite{shaaban2021enhanced}. Due to the low penetration, mobility is also a challenge~\cite{tang2022energy}. On the other hand, lasers offer the advantage of transmitting high power over long distances through collimation, with the risk that safety cannot be guaranteed~\cite{fon2022energy}.

As a new type of free-space optics, the resonant beam (also known as intra-cavity laser) wireless transmission technology, which uses the spatially separated laser resonant cavity as transmitter and receiver, and the resonant beam to transmit energy and information in free space, was proposed in~\cite{liu2016charging}. Collimation of the infrared resonant beam allows energy to be transmitted over long distances with high power~\cite{Zhang2019dis}. Furthermore, the inclusion of retro-reflectors in the resonant beam system offers the ability to reflect incident beam back in its original direction. By incorporating retro-reflectors, such as cat's eye retro-reflectors and corner cube retro-reflectors, at the transmitter and receiver, the system can achieve self-aligned energy transmission without the need for tracking, even during movement~\cite{liu2022mobile}. Moreover, to ensure radiation safety in free space, a study in~\cite{fang2021safety} revealed that the resonant beam can be automatically interrupted if an external object obstructs the transfer channel, which is achieved through the suppression of gain amplification in the transmitter by the presence of the external object.

Therefore, long-range, mobile, human-safe, high-power energy, and high-rate information transfer can be achieved in the resonant beam-based SLIPT system. However, it is evident from the aforementioned examples and studies~\cite{liu2016charging},~\cite{fang2021safety} that the resonant beam can only be transmitted in a line-of-sight (LOS) channel. Meanwhile, non-line-of-sight (NLOS) transmission is a fundamental challenge in FSO transmission, including VL and laser technologies~\cite{najafi2021intelligent},~\cite{ajam2022modeling}. Common methods employed to achieve NLOS transmission involve constructing reconfigurable beam-shaping systems, utilizing diffuse reflections from scattering surfaces (such as walls or ceilings), and installing reflectors~\cite{elgala2011indoor}. However, these strategies possess drawbacks such as complex management, compromised reception quality, and high costs.

Recently, reconfigurable intelligent surfaces (RIS), also known as intelligent reflecting surfaces (IRS), have garnered significant research interest as a promising technology for manipulating the wireless propagation environment of electromagnetic waves~\cite{wu2019towards}. A RIS comprises an array of passive units capable of independently modifying the incident signal with flexibility in parameters such as amplitude, frequency, phase, or polarization~\cite{li2021joint}. By adjusting the parameters of incident electromagnetic waves, the RIS, which is typically mounted on ceilings or walls, can effectively reconstruct the wireless transmission channel between the transmitter and receiver without power consumption and additional noise introduction. For NLOS transmission, RIS can utilize wave reflection to establish a virtual LOS link, effectively overcoming the dead zone.

\begin{figure}[t]
\vspace{-0.8em}
\setlength{\abovecaptionskip}{-5pt}
\setlength{\belowcaptionskip}{-10pt}
    \centering
   \includegraphics[scale=0.48]{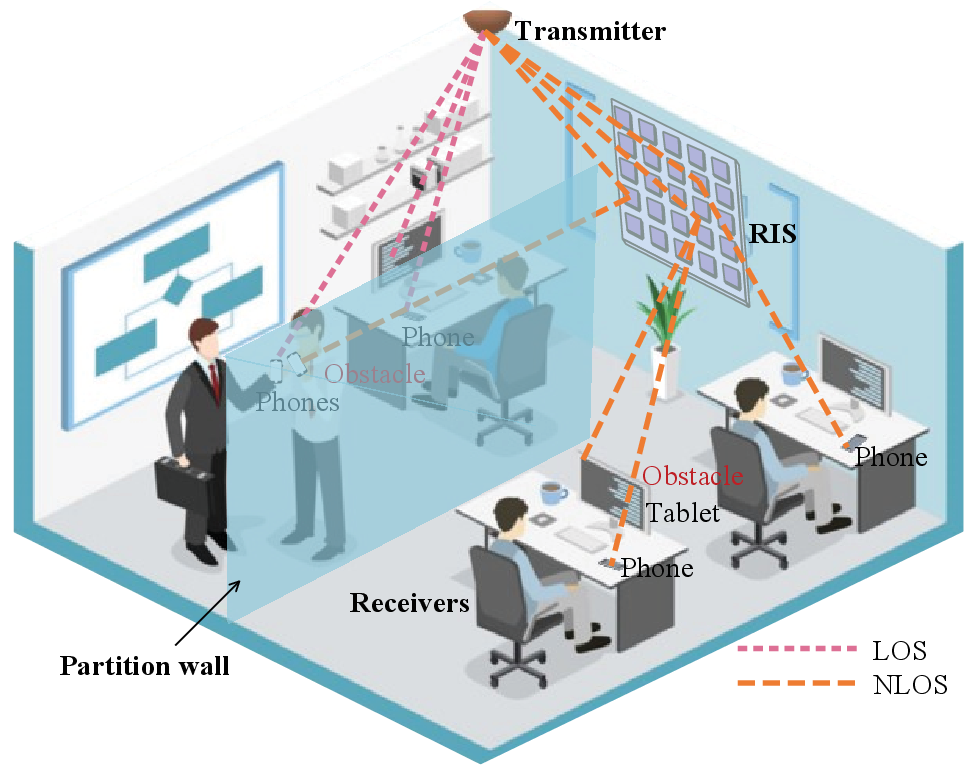}
    \caption{The exemplary application scene of RIS-assisted resonant beam S-SWIPT system.}
    \label{fig:swiptscence}
    \vspace{-1.5em}
\end{figure}

In this paper, we propose a RIS-assisted resonant beam simultaneous wireless information and power transfer (SWIPT) system to achieve highly sensitive detection of LOS channel intrusion, high-power charging, and high-rate communication. Figure~\ref{fig:swiptscence} illustrates an example application of the system, where a transmitter is installed on the ceiling, an RIS is positioned on the wall, and receivers are integrated into the terminals. By adjusting the phase of the resonant beam, the RIS facilitates NLOS transmission in scenarios where the LOS channels are obstructed by external objects, such as human hands or walls.

To analyze channel state information (CSI) in the RIS-assisted resonant beam SWIPT system, we first construct the optical field propagation model between the transceivers by applying the Rayleigh-Sommerfeld diffraction theory of electromagnetic waves, which can provide the optical field distribution on the transfer apertures, enabling precise detection of the obstructed LOS channel through the analysis of changes in the optical field distribution. Subsequently, the theoretical evaluation of the transmission performance is conducted using the transfer efficiency derived from the channel model, output power model, and power-splitting (PS) method. Finally, the performance is numerically analyzed, considering factors such as channel distance, RIS placement, and receiver movement. The contributions of this work can be summarized as follows.
\begin{itemize}
  \item To realize LOS obstacle detection and NLOS high-power transmission in free space, we propose a system architecture for RIS-assisted resonant beam SWIPT, in which a narrow optical beam performs as a free-space carrier and a spatially separated resonator serves as the transmitter and receiver.
  \item Drawing upon the optical field propagation theory, we develop a channel model for the transmission of resonant beams with RIS assistance, which enables the theoretical derivation of the optical field transformation in the transmitter, free space, and moving receiver. Consequently, the steady optical field distribution can be obtained, which forms the foundation for evaluating LOS invasion detection and assessing transmission performance.
  \item We propose a computational approach for performance evaluation in RIS-assisted resonant beam SWIPT systems, which combines an output power model and the power-splitting (PS) method. Employing the approach, we can accurately calculate the transmission efficiency, determine the charging power required for long-endurance operations, and obtain the channel capacity for communication purposes.
  \item We conduct a quantitative evaluation in the RIS-assisted resonant beam SWIPT system considering various parameters, such as RIS placement, channel distance, and receiver mobility. The results highlight the system's proficiency in accurately detecting the invasion of external objects by analyzing optical intensity variations. Moreover, the system can achieve about $4\rm{W}$ charging power, above $11\rm{bps/Hz}$ spectral efficiency, and approximately $55\%$ energy efficiency within a transmission distance of $10\rm{m}$ and specific motion angles.
\end{itemize}

\vspace{-1.0em}
\subsection{Related Works}\label{}
In this subsection, we provide a summary of recent research efforts focusing on RIS-assisted FSO transmission. To overcome the fundamental limitation of the FSO system, the research of RIS-assisted FSO communication is gradually attracting the attention of academicians.

\cite{jamali2021intelligent} offers a comprehensive overview of optical RIS, covering its realization, basic operation principles, advantages, and limitations. To enable the reflection of incident beams in any desired direction within a Gaussian laser beam, \cite{najafi2021intelligent} designs a phase-shift distribution across the IRS and demonstrates the existence of an equivalent mirror-assisted FSO system, whose reflected electric fields are identical to those of the corresponding IRS-assisted systems. For signal-aperture FSO systems without a direct link between the source and destination, \cite{chapala2021unified} provides exact closed-form expressions to evaluate the performance of RIS-empowered FSO systems under various channel impairments, including atmospheric turbulence, pointing errors, and different weather conditions. For MIMO FSO communication, an analytical channel model for point-to-point IRS-assisted FSO systems based on the Huygens-Fresnel principle is developed in~\cite{ajam2022modeling}, and time division (TD), IRS-division (IRSD), and IRS homogenization (IRSH) protocols are proposed for allowing the sharing of the optical IRS by multiple FOS links. To enhance control and management of the transmission environment,~\cite{ndjiongue2022digital} proposes the use of digital metasurfaces as the basis for digital RIS (DRIS), which employs software programming to regulate various processes in RIS elements, including scattering, absorption, reflection, diffraction, and refraction. In addition, \cite{sun2021intelligent} proposes a system implementation of RIS-aided VLC in both signal model and hardware architectures, and illustrates the advantages and potential applications of IRS-assisted VLC, including signal coverage expansion, relaxation of illumination requirements, and signal power enhancement.
\vspace{-1.0em}
\subsection{Organizations}\label{}
In the rest of this paper, Section II provides an overview of the system structure, threshold condition, obstruction detection, and the characteristics of the optical RIS. In Section III, we establish a free-space transmission channel model that incorporates the optical field propagation theory and the field transformation in the transmitter, free space, and receiver. Section IV introduces an analytical model for the output power, considering both electric power conversion and information reception. The performance of the detection and SWIPT is evaluated in Section V, considering different parameters. Finally, in Section VI, we conclude the work and discuss potential directions for future research.

\section{System Overview}\label{Section2}

\begin{figure*}[t]
\vspace{-0.8em}
\setlength{\abovecaptionskip}{-5pt}
\setlength{\belowcaptionskip}{-10pt}
    \centering
    \includegraphics[scale=0.54]{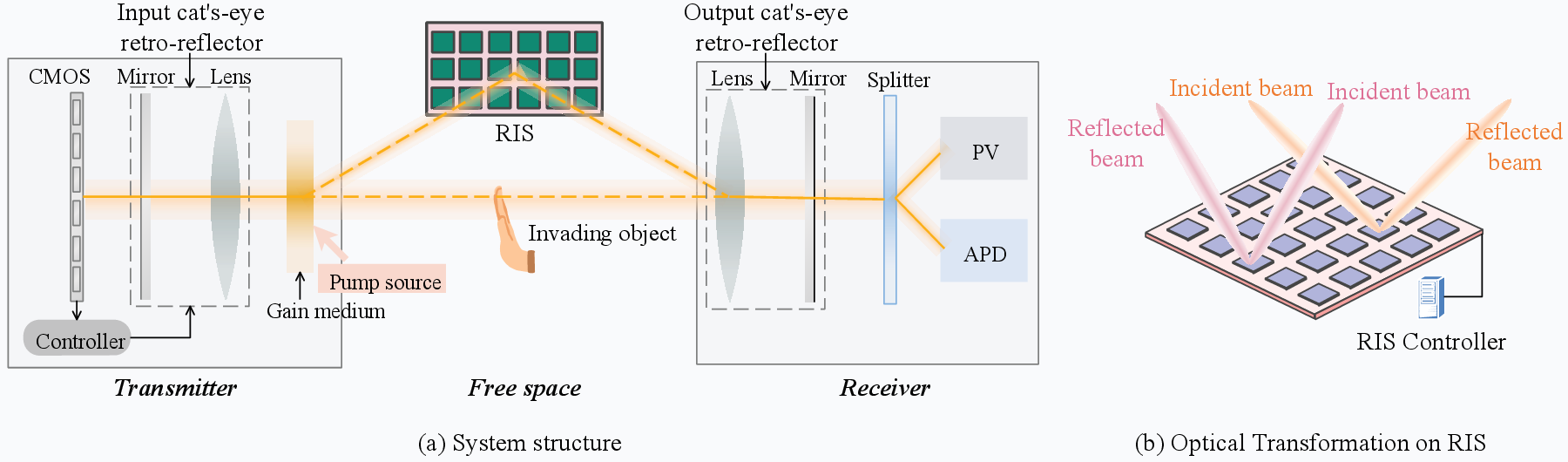}
    \caption{The structure of RIS-assisted resonant beam SWIPT system: (a) System structure; (b) Reflective properties of RIS.}
    \label{fig:system}
    \vspace{-2.0em}
\end{figure*}
\vspace{-1.0em}
\subsection{System Structure}\label{}
An example system structure of RIS-assisted resonant beam S-SWIPT is shown in Fig.~\ref{fig:system}(a). The transmission, reception, and conversion of energy and information are facilitated by spatially separated transmitter and receiver modules. In the transmitter, the power source provides the input power necessary for inducing population inversion and energy level transition of photons in the gain medium to stimulate the resonant beam. The system incorporates a complementary metal-oxide-semiconductor (CMOS) sensor and a controller as its detection components. The CMOS sensor is responsible for imaging the field distribution on the input reflector while the controller is utilized to identify variations in the field distribution, enabling the regulation of power allocation between LOS and NLOS channels.

The output cat's eye reflector functions as a beam splitter for the receiver, with a reflectivity denoted as $R_{\rm{out}}$. A portion of the resonant beam is reflected towards the gain medium to achieve amplification, while the remaining portion is transmitted for energy conversion and information reception. Subsequently, a beam splitter is employed to divide the emitted beam into a power stream and an information stream, with a predetermined ratio of $\gamma$. The power stream is harnessed by a photovoltaic (PV) panel to generate electricity, while the information stream is collected by an avalanche photodiode (APD).

In the system shown in Fig.~\ref{fig:system}(a), the energy transfer is conducted in three steps: power pumping in the transmitter, power transfer via free space, and power output in the receiver. The free space transmission encompasses various scenarios, including fluctuations in the transmission distance and the presence of external objects, resulting in diverse CSI. Specifically, LOS transmission solely involves the transmitter-receiver (T-R) link, while NLOS transmission encompasses both the transmitter-RIS (T-RIS) link and RIS-receiver (RIS-R) link.


\vspace{-10pt}
\subsection{Threshold Power in Resonator}\label{}

\begin{figure}[t]
\vspace{-0.5em}
\setlength{\abovecaptionskip}{0pt}
\setlength{\belowcaptionskip}{-5pt}
    \centering
    \subfigcapskip=0pt
    \subfigure[System parameters and the changes of intensity $I$ in a round trip.]{
    \includegraphics[width=3.6in]{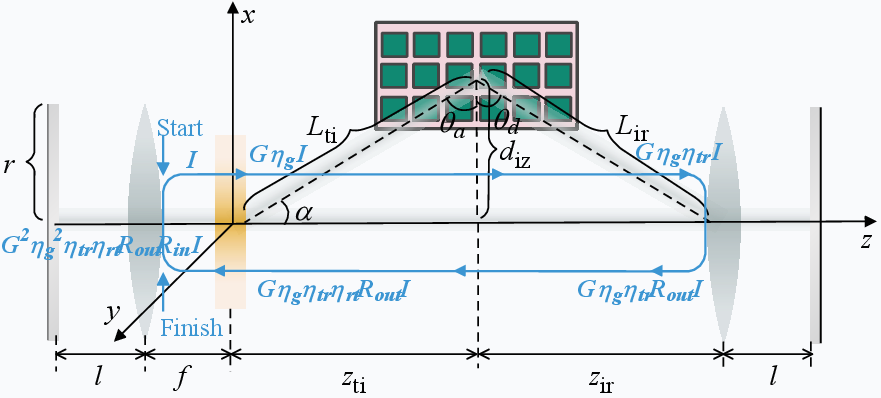}
        \label{fig:aberrationa}}
    \subfigure[Channel functions and field distributions in a round-trip transmission.]{\includegraphics[width=3.6in]{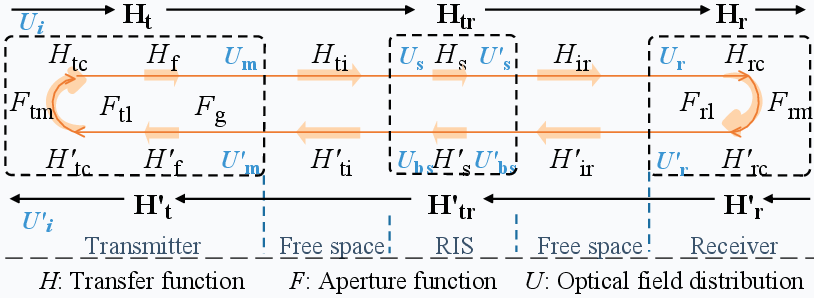}
    \label{fig:aberrationb}}
    \caption{The parameters and light intensity changes of system and the transfer functions in transmitter, free-space, and receiver.}
    \label{fig:parameters}
    \vspace{-1.0em}
\end{figure}

In the transmitter,  the pump source initially provides the input electric power $P_{in}$ to simulate the gain medium, thereby enabling the emission of the intra-cavity beam. As shown in Fig.~\ref{fig:aberrationa}, the initial intensity at the input cat's eye retro-reflector is assumed as $I$, the steady state condition for the intensity in a round-trip transmission reads~\cite{hodgson2005laser}:
\begin{equation}\label{eq:R-2}
\setlength{\abovedisplayskip}{3pt}
\setlength{\belowdisplayskip}{3pt}
    \begin{aligned}
    G^2R_{\rm{in}}R_{\rm{out}}\eta_{\rm{tr}}\eta_{\rm{rt}}\eta_{\rm{g}}^2=1,
    \end{aligned}
\end{equation}
where $G$ represents the light intensity gain factor during each transit through the medium, which is a function of the intensity $I$: $G=exp \left[g_0 \ell /(1+I/I_{\rm{g}})\right]$ with the small-signal gain $g_0 \ell$ and the saturation intensity $I_{\rm{g}}$, representing the peak intensity in gain medium if the light repeatedly passes through the same medium and can be determined by both parameters on the material properties and atomic constants. $R_{\rm{in}}$ and $R_{\rm{out}}$ are the reflectivity of input and output reflectors with $0-100\%$. Additionally, $\eta_{\rm{tr}}$ and $\eta_{\rm{rt}}$ denote the forward and backward transfer efficiencies between the transmitter and receiver in free space, while $\eta_{\rm{g}}$ signifies the transmission efficiency of one propagation direction within the gain medium.

Given that both the forward and backward traveling waves contribute to the beam gain, the average intensity for a single propagation direction can be obtained from~\eqref{eq:R-2} as
\begin{equation}\label{eq:R-3}
\setlength{\abovedisplayskip}{3pt}
\setlength{\belowdisplayskip}{3pt}
    \begin{aligned}
    I=\frac{I_{\rm{g}}}{2}\left[\frac{g_0\ell}{|\ln(\sqrt{R_{\rm{in}}R_{\rm{out}}\eta_{\rm{t}}}\eta_{\rm{g}})|}-1\right],
    \end{aligned}
\end{equation}
where $\eta_{\rm{t}}$ is a round-trip transmission efficiency and can be calculated as $\eta_{\rm{t}}=\eta_{\rm{tr}}\eta_{\rm{rt}}$. In addition, the small-signal gain, which increases with the pump power, determines the amplification of light whose intensity is small compared to the saturation intensity~\cite{hodgson2005laser}:
\begin{equation}\label{eq:R-5}
\setlength{\abovedisplayskip}{3pt}
\setlength{\belowdisplayskip}{3pt}
    \begin{aligned}
     g_0 \ell=\frac{\eta_{\rm{excit}} P_{\rm{in}}}{A_{\rm{g}} I_{\rm{g}}},
    \end{aligned}
    \end{equation}
where $P_{\rm{in}}$ is the input power, $\eta_{\rm{excit}}$ being the excitation efficiency relating the power in the upper laser level to the input power. $A_{\rm{g}}$ is the cross-sectional area of gain medium.

Equation~\eqref{eq:R-3} indicates that the small-signal gain must exceed a specific threshold value to attain laser oscillation within the resonator. Hence, the input power in the system must exceed the threshold power of the resonator, denoted as $P_{\rm{th}}$ ($P_{\rm{in}}>P_{\rm{th}}$). Here, $P_{\rm{th}}$ represents the minimum input electric power necessary to initiate the emission of light from the gain medium. Based on~\eqref{eq:R-3} and~\eqref{eq:R-5}, the threshold power reads:
\begin{equation}\label{eq:3-2}
\setlength{\abovedisplayskip}{3pt}
\setlength{\belowdisplayskip}{3pt}
\begin{aligned}
    P_{\rm{th}} = \frac{A_{\rm{g}} I_{\rm{g}}}{\eta_{\rm{excit}}}|\ln(\sqrt{R_{\rm{in}}R_{\rm{out}}\eta_{\rm{t}}}\eta_{\rm{g}})|.
\end{aligned}
\end{equation}

Finally, as shown in Fig.~\ref{fig:aberrationb}, we assume that the field distribution resonant beam in the transmitter is $\mathbf{U_{\rm{i}}}$ with the pump power satisfying the threshold power, the field distribution after a round trip-transmission reads:
\begin{equation}\label{eq:3-2-revision}
\setlength{\abovedisplayskip}{3pt}
\setlength{\belowdisplayskip}{3pt}
\begin{aligned}
    \mathbf{U_{\rm{i}}'} &= \mathbf{H_{\rm{round}}}\mathbf{U_{\rm{i}}} \\
    &= \mathbf{H_{\rm{t}}}\mathbf{H_{\rm{tr}}}\mathbf{H_{\rm{r}}}\mathbf{H'_{\rm{r}}}\mathbf{H'_{\rm{tr}}}\mathbf{H'_{\rm{t}}}U_{\rm{i}},
\end{aligned}
\end{equation}
where $\mathbf{H_{\rm{round}}}$ represents the transfer function of a round-trip transmission, $\mathbf{H_{\rm{t}}}$, $\mathbf{H_{\rm{tr}}}$, and $\mathbf{H_{\rm{r}}}$ are the forward transfer functions in transmitter, free space, and receiver, while $\mathbf{H'}$ denotes the backward transfer function of each stage. Subsequently, the transfer efficiency and output power can be derived based on the optical field distribution.

\vspace{-5pt}
\subsection{Obstruction Detection Property}\label{}
As shown in Figs.~\ref{fig:swiptscence} and~\ref{fig:system}, owing to the spatially separated cavity configuration, the transmission channel can be invaded by external objects at any time. Such invasions can obstruct the transmission aperture, consequently influencing the resonant beam's ability to generate stable optical wave oscillation modes with the resonator~\cite{fox1968computation}.

\begin{figure}[t]
\setlength{\abovecaptionskip}{-5pt}
\setlength{\belowcaptionskip}{-10pt}
    \centering
    \includegraphics[scale=0.55]{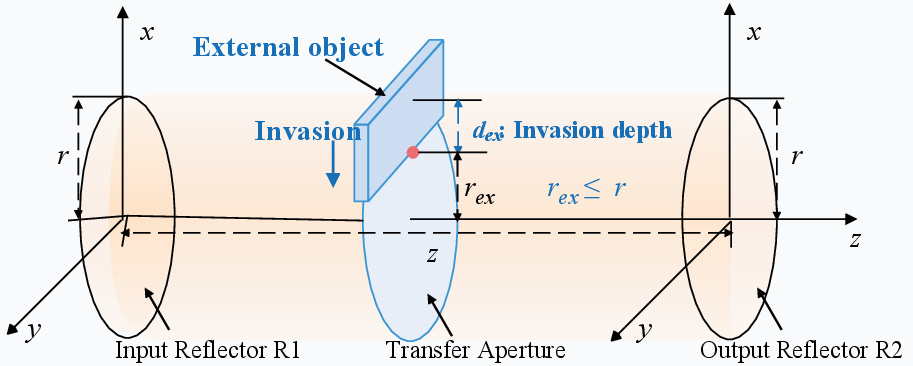}
    \caption{The example diagram of external object invasion into free space transmission channel.}
    \label{fig:invasionape}
    \vspace{-2.0em}
\end{figure}

Thus, an external object invading the transmission channel is equivalent to inserting a new transfer aperture with a reduced radius into the free space, as shown in Fig.~\ref{fig:invasionape}. The radius of the transfer aperture $r_{\rm{ex}}$ diminishes from the original radius $r$ as the invasion depth $d_{\rm{ex}}$ increases. Owing to the resonant beam obstructed by external objects exhibits reduced energy reflection, leading to gain suppression and continuous alterations in the optical field distribution on the reflectors. These variations can be detected sensitively by CMOS. The attenuation of the energy field due to invasion leads to a decline in transfer efficiency and output power. As the invasion depth increases, the output power eventually diminishes to zero.


\vspace{-15pt}
\subsection{Reflective Properties of RIS}
RIS is an engineered surface with adjustable electromagnetic properties, achieved through the utilization of metamaterial technology. As depicted in Fig.~\ref{fig:system}(b), a RIS comprises two primary components: the metamaterial surface and the control module. The metamaterial surface is composed of a multitude of low-cost and passive array elements, which can be equivalently characterized as RLC circuits. By employing the programmable control module, the dynamic manipulation of electromagnetic waves, including phase shift and amplitude adjustment, can be achieved through the RIS~\cite{wu2019towards}.

For an element of the RIS, we assume that the incident signal is $S_{\rm{i}}$, then the reflected signal $S_{\rm{r}}$ can be expressed as~\cite{wu2019towards}
\begin{equation}\label{eq:2-1}
    \begin{aligned}
    \left\{\begin{array}{l}
         S_{\rm{r}}= H_{s} S_{\rm{i}}\\
        H_{s} = \beta \exp{(j\theta)}
    \end{array}\right..
    \end{aligned}
    \end{equation}
where $H_{s}$ is the transformation coefficient of the element including the phase shift and amplitude reflection. $\beta \in [0,1]$ and $\theta \in [0,2\pi)$ are the amplitude reflection coefficient and the phase shift argument of the element, respectively. $j$ is the imaginary unit.



\vspace{-8pt}
\section{Transmission Channel Model}
The resonant beam is pumped by the gain medium and then transmitted through free space to the receiver for power output. According to the Huygens-Fresnel principle, any unobstructed point on the wavefront can be regarded as a secondary wave source, generating spherical sub-waves. The light oscillations at any point in the subsequent space are a coherent superposition of these sub-waves~\cite{fox1968computation}. Therefore, to further characterize the changes in CSI of each channel, we offer the electromagnetic wave propagation and optical field distribution of each optical plane to measure the energy transfer in the RIS-assisted resonant beam SWIPT.

\subsection{Field Transformation in Transmitter}
In the transmitter, the resonant beam will pass through input reflector and gain medium in turn. Assume that the initial complex optical field distribution of incident beam is $U_{\rm{m}}$ ($U_{\rm{m}} \in \mathbb{C}^{M\times N}$, where $\mathbb{C}^{M\times N}$ refers to the space of $M\times N$ complex matrices, $M\times N$ represents the number of secondary wave sources), the output field distribution can be calculated as
\begin{equation}\label{eq:3-3-field}
\setlength{\abovedisplayskip}{3pt}
\setlength{\belowdisplayskip}{3pt}
\begin{aligned}
   U_{\rm{m}}'=\mathscr{F}^{-1}\left\{\mathscr{F}\left\{\mathbf{U_{\rm{m}}}\right\}\mathbf{H_{\rm{T}}}\right\},
   \end{aligned}
\end{equation}
where $\mathscr{F}$ and $\mathscr{F}^{-1}$ are two-dimension Fourier transform and inverse Fourier transform. $\mathbf{H_{\rm{T}}}$ represent the channel function of transmitter and can be  expressed as (Fig.~\ref{fig:parameters}(b))
\begin{equation}\label{eq:3-3}
\setlength{\abovedisplayskip}{3pt}
\setlength{\belowdisplayskip}{3pt}
\begin{aligned}
   \mathbf{H_{\rm{T}}} = \mathbf{H_{\rm{t}}}\mathbf{H'_{\rm{t}}}= F_{\rm{g}} H_{\rm{f}}' F_{\rm{tl}} H_{\rm{tc}}' F_{\rm{tm}} H_{\rm{tc}} F_{\rm{tl}}  H_{\rm{f}}' F_{\rm{g}},
\end{aligned}
\end{equation}
where $F_{\rm{g}}$, $F_{\rm{tl}}$, and $F_{\rm{tm}}$ denote the aperture functions of the gain medium, focal lens and reflecting mirror in the cat's eye reflector. $H_{\rm{f}}$ and $H_{\rm{tc}}$ represent the channel function from the focal lens to gain medium, and from the reflecting mirror to the focal lens. Meanwhile, $\rm{H}'$ is the channel function for the above transmission process in the reverse direction.

The channel function can be calculated based on Rayleigh-Sommerfeld diffraction formula as
\begin{equation}\label{eq:3-7}
\begin{aligned}
    H\left(f_x, f_y\right)&=\mathscr{F}\left\{\frac{z}{j \lambda} \frac{\exp \left(j k \sqrt{x^2+y^2+z^2}\right)}{x^2+y^2+z^2}\right\}
    &=\exp \left\{j k z \sqrt{1-\left(\lambda f_x\right)^2-\left(\lambda f_y\right)^2}\right\},
\end{aligned}
\end{equation}
where $(x,y)$ represents the horizontal and vertical coordinates in the source plane, and $(f_x, f_y)$ represents the spatial frequency coordinate in the transform domain. $\lambda$ is the wavelength, $k=2\pi/\lambda$ is the wavenumber. $z$ denotes the distance between the transfer apertures. The aperture function is limited to the size of transfer aperture, reads
\begin{equation}\label{eq:3-8}
\begin{aligned}
    F(x,y) = \left\{\begin{array}{l}
\exp \left[\frac{-j \pi}{\lambda f}\left(x^2+y^2\right)\right], \ \  x^2+y^2 \leq r^2 \\
0, \quad x^2+y^2>r^2
\end{array}\right.,
\end{aligned}
\end{equation}
where $r$ is the radius of the transfer aperture. If the wave source point $(x,y)$ is inside the aperture, the aperture function is related to the wavelength $\lambda$ and focal length $f$ of the aperture, whereas $F(x,y) = 0$ if the point $(x,y)$ is outside the aperture. Additionally, the aperture function of external object invasion plane $F_{\rm{ex}}$ also depends on the invasion depth $d_{\rm{ex}}$ shown in Fig.~\ref{fig:invasionape}. For example, if the external object invades in the negative direction of $x$-axis, the aperture function for invasion plane can be expressed as
\vspace{-0.5em}
\begin{equation}\label{eq:3-8-revision}
\setlength{\belowdisplayskip}{-1.0em}
\begin{aligned}
    F_{\rm{ex}} = \left\{\begin{array}{l}
\exp \left[\frac{-j \pi}{\lambda f}\left(x^2+y^2\right)\right], \ \  x^2+y^2 \leq r^2 \ \&\& \ x\leq r-d_{\rm{ex}}\\
0, \quad x^2+y^2>r^2 \ || \ x > r-d_{\rm{ex}}
\end{array}\right..
\end{aligned}
\end{equation}

\subsection{RIS-Assisted NLOS Transmission Model}
For RIS-assisted NLOS transmission, we define the transfer functions of T-RIS, RIS, RIS-R channel as $H_{\rm{ti}}$, $H_{\rm{s}}$, and $H_{\rm{ir}}$. Then, the channel function in free space $\mathbf{H_{\rm{F}}}$ is given by (Fig.~\ref{fig:parameters}(b))
\begin{equation}\label{eq:3-9}
\setlength{\abovedisplayskip}{3pt}
\setlength{\belowdisplayskip}{3pt}
\begin{aligned}
   \mathbf{H_{\rm{F}}}=\mathbf{H_{\rm{f}}}\mathbf{H'_{\rm{f}}} = H_{\rm{ti}}H_{\rm{s}}H _{\rm{ir}}H' _{\rm{ir}}H'_{\rm{s}}H'_{\rm{ti}}.
\end{aligned}
\end{equation}
Similarly, the transfer function in the opposite direction of transmission is denoted as $H'$. The calculations for each transmission channel are provided below.

\subsubsection{From transmitter to RIS (T-RIS channel)}

\

Based on the complex optical field distribution output from the transmitter, denoted as $U_{\rm{m}}'$, the incident beam field distribution in RIS $U_{\rm{s}}$ can be illustrated as
\begin{equation}\label{eq:3-10}
\setlength{\abovedisplayskip}{3pt}
\setlength{\belowdisplayskip}{3pt}
\begin{aligned}
    U_{\rm{s}}(x_{\rm{s}},y_{\rm{s}}) = \mathscr{F}^{-1}\left\{\mathscr{F}\left\{U_{\rm{m}}'(x_{\rm{m}}, y_{\rm{m}})\right\} H_{\rm{ti}}(f_{x\rm{m}}, f_{y\rm{m}}, L_{\rm{ti}})\right\} ,
\end{aligned}
\end{equation}
with
\begin{equation}\label{eq:3-11}
\setlength{\abovedisplayskip}{3pt}
\setlength{\belowdisplayskip}{3pt}
\begin{aligned}
    H_{\rm{ti}}(f_{x\rm{m}}, f_{y\rm{m}}, L_{\rm{ti}}) = \exp \left\{j k L_{\rm{ti}} \sqrt{1-\left(\lambda f_{x\rm{m}}\right)^2-\left(\lambda f_{y\rm{m}}\right)^2}\right\},
\end{aligned}
\end{equation}
where $(x_{\rm{m}},y_{\rm{m}})$ and $(x_{\rm{s}},y_{\rm{s}})$ represent the elements on the surface of the gain medium and RIS, respectively. Additionally, $L_{\rm{ti}}$ is the channel distance between the transmitter and RIS,
\begin{equation}\label{eq:3-12}
\setlength{\abovedisplayskip}{3pt}
\setlength{\belowdisplayskip}{3pt}
\begin{aligned}
    L_{\rm{ti}} = \frac{z_{\rm{ti}}}{\cos(\alpha)} = \frac{z_{\rm{ti}}}{\sin[\arctan(\frac{d_{\rm{iz}}}{z_{\rm{ti}}})]},
\end{aligned}
\end{equation}
where $\alpha$ is the angle between $z$-axis and the line between transmitter and RIS, $z_{\rm{ti}}$ is the distance between transmitter and RIS in $z$-axis, and $d_{\rm{iz}}$ is the vertical height of RIS, as shown in Fig~\ref{fig:aberrationb}.

\subsubsection{Field Transformation in RIS}
\
\newline
\indent Both the amplitude and phase of the incident field distribution $U_{\rm{s}}$ can be altered by RIS. Consequently, the emitted optical field distribution $U_{\rm{s}}'$ can be derived as
\begin{equation}\label{eq:3-13}
\setlength{\abovedisplayskip}{3pt}
\setlength{\belowdisplayskip}{3pt}
\begin{aligned}
    U_{\rm{s}}' = U_{\rm{s}} H_{\rm{s}} = U_{\rm{s}}\cdot\beta exp(j\theta),
\end{aligned}
\end{equation}
where the value of $\beta$ is set to $1$ to ensure maximum energy transfer efficiency. As shown in Fig.~\ref{fig:parameters}(a), the angle $\theta$ represents the phase shift angle of the RIS, which is determined by the angle-of-arrival (AoA) from the transmitter to the RIS $\theta_{\rm{a}}$, and the angle-of-departure (AoD) from the RIS to the receiver $\theta_{\rm{d}}$. The angles can be calculated as:
\begin{equation}\label{eq:3-14}
\setlength{\abovedisplayskip}{3pt}
\setlength{\belowdisplayskip}{3pt}
    \begin{aligned}
         \theta_{\rm{a}}= \arctan\left(\frac{z_{\rm{ti}}}{d_{\rm{iz}}}\right), \ \         \theta_{\rm{d}}= \arctan\left(\frac{z_{\rm{ir}}}{d_{\rm{iz}}}\right)
    \end{aligned}
    \end{equation}
where $z_{\rm{ti}}$ and $z_{\rm{ir}}$ represent the distances between the transmitter and RIS, and between the transmitter and RIS along the $z$-axis, as depicted in Fig.~\ref{fig:aberrationa}. Additionally, $d_{\rm{iz}}$ denotes the distance from the RIS to the $z$-axis.


\subsubsection{From RIS to receiver (RIS-R channel)}

\

The optical field $U_{\rm{s}}'$ emitted by the RIS propagates through free space towards the receiver. Upon reaching the receiver, the field distribution undergoes various transformations, including rotation and translation. Firstly, the spectrum of the received field distribution $U_{\rm{r}}$ in the receiver (i.e. the front surface of the focusing mirror in the output cat's-eye reflector) is defined by~\cite{matsushima2003fast}
\begin{equation}\label{eq:3-15}
\setlength{\abovedisplayskip}{3pt}
\setlength{\belowdisplayskip}{3pt}
\begin{aligned}
    G(f_{x\rm{r}}, f_{y\rm{r}}) = \mathscr{F}\left\{U_{\rm{s}}'(x_{\rm{s}}, y_{\rm{s}})\right\} H_{ir}(f_{x\rm{s}}, f_{y\rm{s}}, L_{ir}),
\end{aligned}
\end{equation}
where the channel function is
\begin{equation}\label{eq:3-16}
\setlength{\abovedisplayskip}{3pt}
\setlength{\belowdisplayskip}{3pt}
\begin{aligned}
    H_{ir}(f_{x\rm{s}}, f_{y\rm{s}}, L_{ir}) = \exp \left\{j k L_{\rm{ir}} \sqrt{1-\left(\lambda f_{x\rm{s}}\right)^2-\left(\lambda f_{y\rm{s}}\right)^2}\right\}.
\end{aligned}
\end{equation}
The distance between the RIS and receiver is expressed as
\begin{equation}\label{eq:3-17}
\begin{aligned}
    L_{\rm{ir}} = \frac{z_{\rm{ir}}}{\sin{(\theta_{\rm{d}})}},
\end{aligned}
\end{equation}
where $\theta_{\rm{d}}$ can be calculated by using \eqref{eq:3-14}. The Fourier frequency for $(x, y, z)$ is $[f_{x}, f_{y}, w(f_{x}, f_{y})]$, where $w(f_{x}, f_{y}) = \left(\lambda^{-2}-f_{x}^2-f_{y}^2\right)^{1 / 2}$~\cite{de2005angular}. Finally, the two-dimensional complex amplitude distribution in the output reflector can be depicted as
\begin{equation}\label{eq:3-18}
\setlength{\abovedisplayskip}{3pt}
\setlength{\belowdisplayskip}{-1.0em}
\begin{aligned}
    U_{\rm{r}}(x_{\rm{r}},y_{\rm{r}}) = \mathscr{F}^{-1}\left\{G(f_{x\rm{r}}, f_{y\rm{r}})\right\}.
\end{aligned}
\end{equation}

\subsection{Field Transformation in Receiver}
The movements of receiver, such as translation and rotation, will result in changes to the field distribution on the output reflector. Subsequently, we provide a detailed solution for the transformation of the optical field in the moving receiver.

\subsubsection{Translation of Receiver}

\

%

\begin{figure*}[t]
\vspace{-0.5em}
\setlength{\abovecaptionskip}{0pt}
\setlength{\belowcaptionskip}{-10pt}
\centering
\subfigcapskip=-10pt
\subfigure[Translation along the $xoy$ plane.]{
\begin{minipage}[t]{0.45\linewidth}
\centering
\includegraphics[scale=0.21]{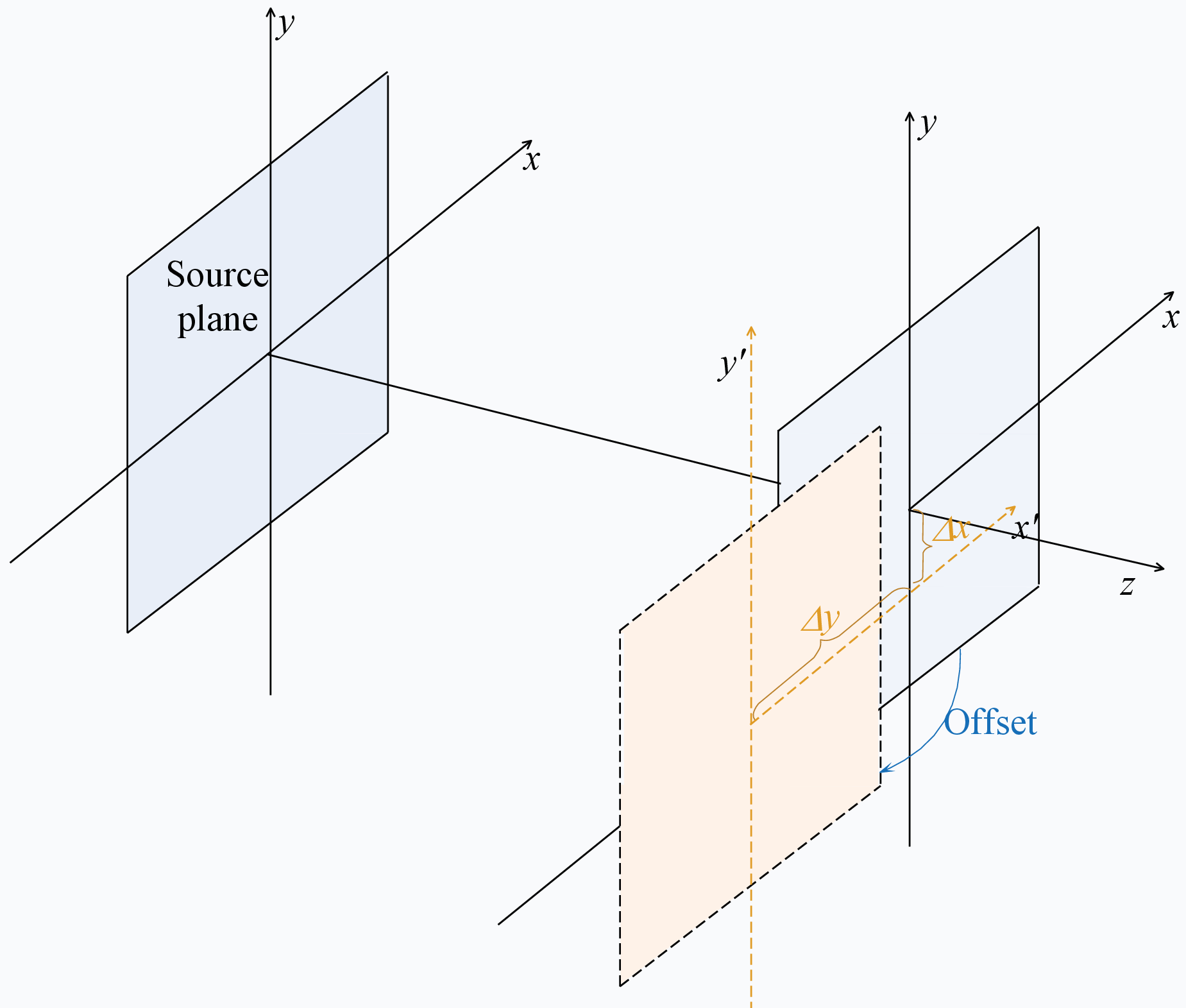}
\label{fig:offset}
\end{minipage}
}
\subfigure[Rotation around $x$, $y$, and $z$ axes.]{
\begin{minipage}[t]{0.45\linewidth}
\centering
\includegraphics[scale=0.22]{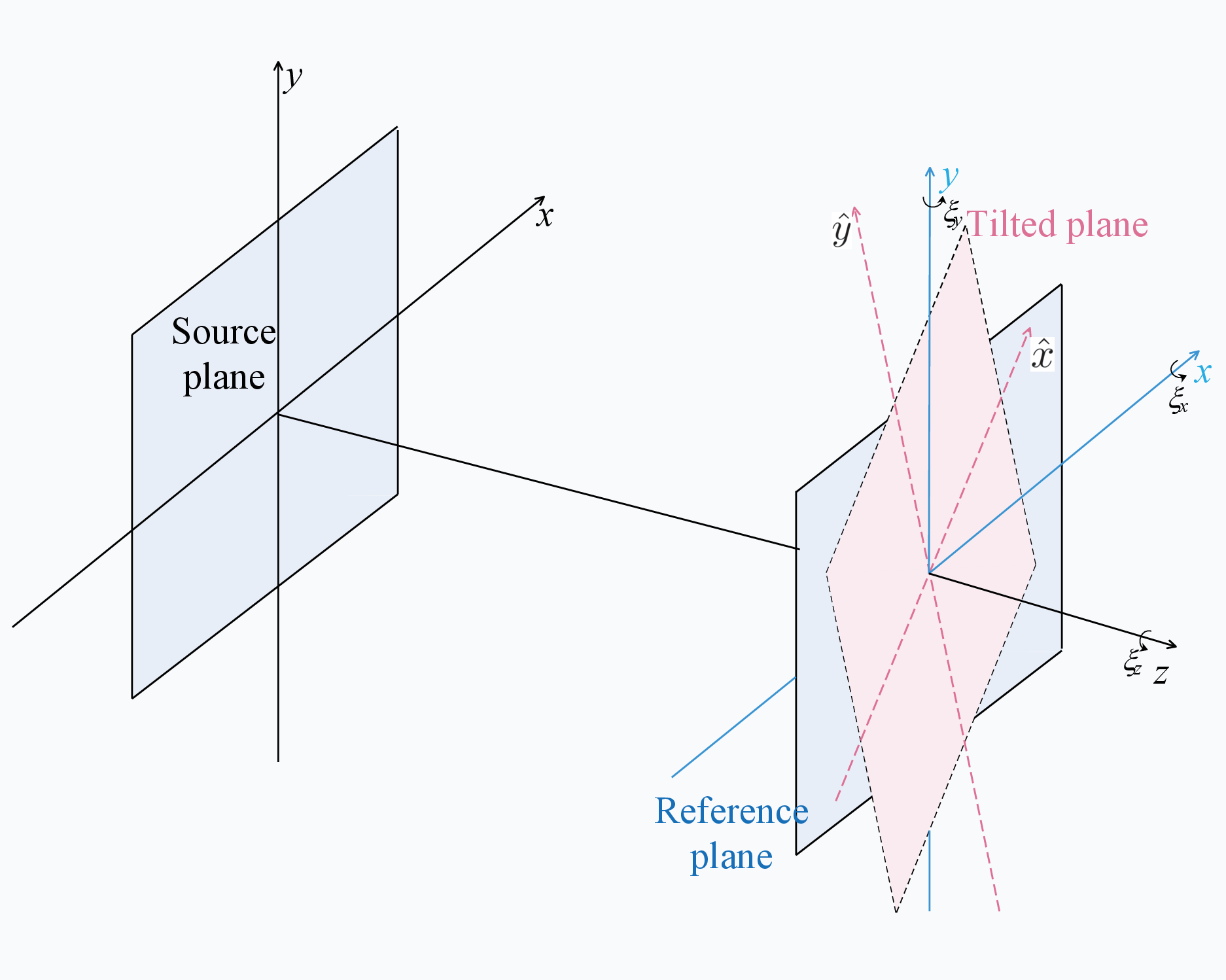}
\label{fig:title}
\end{minipage}
}
\centering
\caption{The schematic diagram of the movement receiver, where $\Delta x$ and $\Delta y$ represent the translation distances, and $\xi_{x}$, $\xi_{y}$, and $\xi_{z}$ denote the rotation angles.}
\label{Fig:movement}
\vspace{-2.0em}
\end{figure*}

As illustrated in Fig.~\ref{fig:offset}, the output reflector becomes offset from the optical axis of the previous plane if the receiver undergoes movement in the $xoy$ plane. The offset coordinate system can be calculated by shifting the previous coordinate system:
    \begin{equation}\label{eq:3-27}
    \setlength{\abovedisplayskip}{3pt}
    \setlength{\belowdisplayskip}{3pt}
    \begin{aligned}
    x^{\prime}=x-x_0, \ \ y^{\prime}=y-y_0, \ \ z^{\prime}=z.
    \end{aligned}
    \end{equation}
where $(x',y',z')$ is the shifted coordinate system, $x_0$ and $y_0$ are the translation distance of receiver. Afterward, based on \eqref{eq:3-18} and \eqref{eq:3-27}, the optical field distribution on the shifted reflector can be expressed as~\cite{delen1998free}
    \begin{equation}\label{eq:3-28}
    \setlength{\abovedisplayskip}{3pt}
    \setlength{\belowdisplayskip}{3pt}
    \begin{aligned}
    U(x{\prime}, y{\prime}) = \mathscr{F}^{-1}\{G_{0}(f_{\rm{x}},f_{\rm{y}})H(f_{\rm{x}}, f_{\rm{x}})\},
    \end{aligned}
    \end{equation}
where $G_{0}(f_{x},f_{y})$ is the spectrum of complex field distribution and can be written as
    \begin{equation}\label{eq:3-29}
    \setlength{\abovedisplayskip}{3pt}
    \setlength{\belowdisplayskip}{3pt}
    \begin{aligned}
     G_{0}(f_{x},f_{y})=\mathscr{F}\{U(x, y)\} \exp \left[i 2 \pi\left(f_x x_0+f_y y_0\right)\right].
    \end{aligned}
    \end{equation}
$U(x, y)$ represents the field distribution of the point $(x,y)$ on the reflector in the coordinate system prior to offsets. Moreover, \eqref{eq:3-29} denotes the shift theorem of Fourier transform theory, which explains that the Fourier transform of a function in a shifted coordinate system equals the Fourier transform of the same function in an unshifted coordinate system, multiplied by a phase that is proportional to the amount of the shift~\cite{goodman1968introduction}.

\subsubsection{Rotation of Receiver}

\


The rotation of the receiver by a certain angle is equivalent to the tilting of the output reflector. For the titled plane, as shown in Fig.~\ref{fig:title}, we define the plane parallel to the incident plane as the reference plane in the coordinate system $(x,y,z)$, and the coordinate of the titled plane is $(\hat{x}, \hat{y}, \hat{z})$. Both coordinate systems share the same origin and are not parallel to each other. Consequently, if the field distribution on the reference plane and the titled plane represent the same wave, the wave vector can be transformed from one coordinate system to the other using coordinate transformation~\cite{matsushima2003fast}.


    \begin{table}[t]
    \vspace{-1.5em}
    \renewcommand\arraystretch{1}
    \setlength{\abovecaptionskip}{-0pt}
    \setlength{\belowcaptionskip}{-2pt}
    \centering
     \caption{Rotation matrices around $x$, $y$, and $z$ axes}
    \begin{tabular}{p{0.6cm}<{\centering}p{4cm}<{\centering}|p{0.6cm}<{\centering}p{4cm}<{\centering}|p{0.6cm}<{\centering}p{4cm}<{\centering}}
    \hline
    \textbf{{\small Symbol}} & \textbf{{\small Matrices}} & \textbf{{\small Symbol}} & \textbf{{\small Matrices}} & \textbf{{\small Symbol}} & \textbf{{\small Matrices}}\\
    \hline
    \bfseries{{\small $\mathbf{R}_x(\xi_{x})$}} & {$\left(\begin{array}{ccc}
    {\small 1} & {\small 0} & {\small 0} \\
    {\small 0} & {\small \cos \xi_x} & {\small \sin \xi_x} \\
    {\small 0} & {\small -\sin \xi_x} & {\small \cos \xi_x}
    \end{array}\right)$
    } &
    \bfseries{{\small $\mathbf{R}_y(\xi_{y})$}} & {$\left(\begin{array}{ccc}
    {\small \cos \xi_y} & {\small 0} & {\small -\sin \xi_y} \\
    {\small 0} & {\small 1} & {\small 0} \\
    {\small \sin \xi_y} & {\small 0} & {\small \cos \xi_y}
    \end{array}\right)$
    } &
    \bfseries{{\small $\mathbf{R}_z(\xi_{z})$}} & {$\left(\begin{array}{ccc}
    {\small \cos \xi_z} & {\small \sin \xi_z} & {\small 0} \\
    {\small -\sin \xi_z} & {\small \cos \xi_z} & {\small 0} \\
    {\small 0} & {\small 0} & {\small 1}
    \end{array}\right)$
    }\\
    \hline
    \label{table1}
    \end{tabular}
    \vspace{-4.0em}
\end{table}

Based on \eqref{eq:3-15} and \eqref{eq:3-16}, the wave vector of field distribution in the reference coordinate is $\mathbf{k}=2\pi[f_x, f_y, w(f_x, f_y)]$, while it is $\mathbf{\hat{k}}=2\pi[\hat{f_x}, \hat{f_y}, \hat{w}(f_x, f_y)]$ with $\hat{w}(\hat{f_x}, \hat{f_y})]=\left(\lambda^{-2}-\hat{f_x}^2-\hat{f_y}^2\right)^{1 / 2}$ in the rotation coordinate~\cite{de2005angular}. Suppose that $\mathbf{M}$ is a rotation matrix, which is used to transform the wave vector in reference coordinate into the titled coordinate and can be defined as a rotation matrix $\mathbf{R}_{\epsilon}(\xi_{\epsilon})$ or the product of several rotation matrices around $x$, $y$, or $z$ axes:
\begin{equation}\label{eq:3-21}
    \setlength{\abovedisplayskip}{1pt}
    \setlength{\belowdisplayskip}{1pt}
    \begin{aligned}
        \mathbf{M} = \mathbf{R}_{\epsilon}(\xi_{\epsilon})\cdots \mathbf{R}_{\tau}(\xi_{\tau}),
    \end{aligned}
\end{equation}
where $\epsilon$ and $\tau$ denote $x$, $y$, or $z$ axes, $\xi_{\epsilon}$ and $\xi_{\tau}$ are the angles of rotation around the axes $\epsilon$ and $\tau$. The rotation matrices around $x$, $y$, and $z$ axes are depicted in Table~\ref{table1}.

Additionally, the relationship between the two wave vectors can be depicted as~\cite{matsushima2008formulation}
\begin{equation}\label{eq:3-22}
    \setlength{\abovedisplayskip}{2pt}
    \setlength{\belowdisplayskip}{2pt}
    \begin{aligned}
        \hat{\mathbf{k}} = \mathbf{M} \mathbf{k}, \ \ \mathbf{k} = \mathbf{M}^{-1} \hat{\mathbf{k}},
    \end{aligned}
    \end{equation}
where $\mathbf{M}^{-1}$ is the inversion transformation matrix of $\mathbf{M}$, reads    \begin{equation}\label{eq:3-23}
\setlength{\abovedisplayskip}{3pt}
    \setlength{\belowdisplayskip}{3pt}
    \begin{aligned}
        \mathbf{M^{-1}}=\left[\begin{array}{lll}
    a_1 & a_2 & a_3 \\
    a_4 & a_5 & a_6 \\
    a_7 & a_8 & a_9
    \end{array}\right].
    \end{aligned}
    \end{equation}
Then, we can associate the Fourier frequencies in the reference coordinate with those in the rotation coordinate
    \begin{equation}\label{eq:3-24}
    \setlength{\abovedisplayskip}{3pt}
    \setlength{\belowdisplayskip}{3pt}
   \begin{aligned}
  \left\{\begin{array}{l}
   f_x=\phi(\hat{f_x}, \hat{f_y})=a_1 \hat{f_x}+a_2 \hat{f_y}+a_3 \hat{w}(\hat{f_x}, \hat{f_y}) \\
    f_y=\psi(\hat{f_x}, \hat{f_y})=a_2 \hat{f_x}+a_5 \hat{f_y}+a_6\hat{w}(\hat{f_x}, \hat{f_y})
    \end{array}\right..
    \end{aligned}
    \end{equation}
Thus, the angular spectrum distribution on the titled plane can be written as
    \begin{equation}\label{eq:3-25}
    \setlength{\abovedisplayskip}{3pt}
    \setlength{\belowdisplayskip}{3pt}
    \begin{aligned}
    \hat{G}(\hat{f_x}, \hat{f_y}) = G(\phi(\hat{f_x}, \hat{f_y}), \psi(\hat{f_x}, \hat{f_y})).
    \end{aligned}
    \end{equation}
Afterward, it appears feasible to compute the complex amplitude within the titled plane through the inverse transformation of the spectrum $\hat{G}(\hat{f_x}, \hat{f_y})$. Nonetheless, due to the nonlinear transformation described in \eqref{eq:3-24}, the total energy of the field is not conserved after the rotational transformation. Thus, a mere inverse Fourier transformation of $\hat{G}(\hat{f_x}, \hat{f_y})$ cannot yield the accurate field distribution result~\cite{matsushima2003fast}. Interpolation is commonly employed as a method to facilitate the transformation from a non-uniform spectrum distribution to a uniformly-sampled distribution. Subsequently, the complex optical field distribution within the titled reflector can be determined by applying the inverse Fourier transformation, while incorporating a Jacobian function to compensate for the nonlinearity of the rotational transformation~\cite{matsushima2003fast}, \cite{matsushima2008formulation},
\begin{equation}\label{eq:3-26}
\setlength{\abovedisplayskip}{3pt}
    \setlength{\belowdisplayskip}{3pt}
\begin{aligned}
    U(x, y)= \mathscr{F}^{-1}\left\{\hat{G}(\hat{f_{x}}, \hat{f_{y}})|J(\hat{f_{x}}, \hat{f_{y}})|\right\},
    \end{aligned}
    \end{equation}
where the Jacobian $J(\hat{f_{x}}, \hat{f_{y}})$ is defined as
\begin{equation}\label{eq:3-26-2}
\setlength{\abovedisplayskip}{3pt}
    \setlength{\belowdisplayskip}{3pt}
    \begin{aligned}
    J(\hat{f_{x}}, \hat{f_{y}}) =& \frac{\partial \phi}{\partial \hat{f_x}} \frac{\partial \psi}{\partial \hat{f_y}}-\frac{\partial \phi}{\partial \hat{f_y}} \frac{\partial \psi}{\partial \hat{f_x}} \\
= & \left(a_2 a_6-a_3 a_5\right) \frac{\hat{f_{x}}}{\hat{w}(\hat{f_{x}}, \hat{f_{y}})} +\left(a_3 a_4-a_1 a_6\right) \frac{\hat{f_{y}}}{\hat{w}(\hat{f_{x}}, \hat{f_{y}})}+\left(a_1 a_5-a_2 a_4\right).
    \end{aligned}
    \end{equation}


Based on the aforementioned analysis, the field distribution on the receiver for a particular rotation angle can be determined by following these steps: a) propagating the field to a reference plane that is parallel to the source surface, b) performing a rotation of the field distribution's spectrum into the titled plane using equations \eqref{eq:3-21}-\eqref{eq:3-25}, and c) calculating the complex optical field distribution within the titled plane using \eqref{eq:3-26}. Next, to achieve backward transmission of the intra-cavity beam, the titled plane needs to be rotated to the reference plane by employing the rotation angles $-\xi_{x}$, $-\xi_{y}$, or $-\xi_{z}$ as prescribed in equations \eqref{eq:3-21} and \eqref{eq:3-22}. Following this rotation, the field propagation can be computed using equations \eqref{eq:3-24}-\eqref{eq:3-26-2}.

\subsubsection{Field Transformation in Output Reflector}

\

Lastly, the optical field distribution $U_r(x_r, y_r)$ on the front surface of the focal lens, resulting from translation or rotation as described in~\eqref{eq:3-28} or~\eqref{eq:3-26}, will propagate through the focal lens, the free space between the focal lens and the mirror, and the mirror. The optical field transformation in the output cat's retro-reflector can be expressed as
\begin{equation}\label{eq:3-3}
\setlength{\abovedisplayskip}{3pt}
    \setlength{\belowdisplayskip}{3pt}
\begin{aligned}
   U_r'(x_r, y_r) &= \mathbf{H_{\rm{t}}}U_r(x_r, y_r) \\
   &= F_{\rm{rl}} H_{\rm{rc}} F_{\rm{rm}} H_{\rm{rc}}'F_{\rm{rl}} U_r(x_r, y_r),
\end{aligned}
\end{equation}
where $F_{\rm{rl}}$ and $F_{\rm{rm}}$ represent the aperture functions of the focal lens and mirror and can be calculated by \eqref{eq:3-8}. $H_{\rm{rc}}$ is the transfer function from the focal lens and mirror, $H_{\rm{rc}}'$ is the function in the opposite propagation direction.

So far, we can quantify the field distribution in a round-trip transmission by employing optical field propagation in transmitter, free space, and receiver. Nevertheless, owing to the occurrence of diffraction loss, the field distribution following a complete round-trip transmission fails to capture the steady-state model. Therefore, it becomes necessary to employ the Fox-Li iteration algorithm, where each iteration represents a round-trip transmission of the resonant beam. Once the iteration count reaches a specific threshold, a self-reproducing mode emerges, wherein the relative field distribution remains constant from one transit to another~\cite{fox1961resonant}.


\vspace{-15pt}
\section{Power Output and Energy Conversion Model}
Based on the steady-state optical field distribution on each plane, the transfer efficiency and output power in the RIS-assisted resonant beam SWIPT system can be obtained.
\vspace{-15pt}
\subsection{Transfer Efficiency}
The transfer efficiency $\eta$ in a round-trip transmission can thus be calculated as follows using the steady-state optical field distribution:
\begin{equation}\label{eq:3-35}
\setlength{\abovedisplayskip}{3pt}
    \setlength{\belowdisplayskip}{3pt}
    \begin{aligned}
     \eta_{n n+1}=\frac{\iint_S\left|U_{n+1}\right|^2 d s}{\iint_S\left|U_n\right|^2 d s},
    \end{aligned}
    \end{equation}
where $|U|^2$ represents the beam intensity distribution, and then its double integral denotes the total energy in the plane. $U_{n}$ and $U_{n+1}$ are the steady-state field distribution at the n-th and (n+1)-th transmission iterations in the same plane. Consequently, the transfer efficiency between different optical planes can be derived by analyzing the energy difference, which is based on the steady-state field distribution on the planes during a round-trip transmission.
\vspace{-15pt}
\subsection{Output Optical Power}
Afterward, based on the characteristics of gain medium and the intra-cavity transmission efficiency, the output beam power of the RIS-assisted resonant beam AS-SWIPT system can be expressed as~\cite{hodgson2005laser}~\cite{hodgson1989optical}
\begin{equation}\label{eq:3-36}
    \begin{aligned}
     \begin{gathered}
P_{\text {out }}=\frac{A_{\rm{b}} I_{\rm{g}}\left(1-R_{\rm{out}}\right) \eta_{\rm{tr}}}{1-R_{\rm{out}} \eta_{\rm{tr}}\eta_{\rm{rt}}+\sqrt{R_{\rm{out}} \eta_{\rm{o}}}\left(\frac{1}{\eta_{\rm{lg}} \eta_{\rm{tr}} \eta_{\rm{g}}}-\eta_{\rm{g}}\right)}
\left(\frac{\eta_{\rm{excit}} P_{\rm{in}}}{A_{\rm{g}} I_{\rm{g}}}-\mid \ln \sqrt{R_{\rm{out}} \eta_{\rm{g}}^2 \eta_{\rm{o}} \mid}\right),
\end{gathered}
    \end{aligned}
    \end{equation}
where $A_{\rm{b}}$ is the cross-sectional area of beam spot on the gain medium. $\eta_{\rm{tr}}$ and $\eta_{\rm{rt}}$ represent the forward and backward transmission efficiencies between the transmitter and receiver. Additionally, $\eta_{\rm{lg}}$ signifies the transmission efficiency from the input reflector's mirror to the gain medium, while $\eta_{\rm{o}}$ corresponds to the overall transmission efficiency of a round-trip.

According to \eqref{eq:3-36}, we can see that the system's output power is influenced by various factors, including the properties of the gain medium, the reflectivity of the output reflector, the transmission efficiency, and the cross-sectional area of the intra-cavity beam. Consequently, the output power is susceptible to changes in the transmission distance, aperture radius, RIS location, and the movement of the output reflector. These variables collectively impact the distribution of the optical field propagation.
\vspace{-15pt}
\subsection{Charging Power and Information Receiving}
In the receiver, the beam splitter divides the output beam into two streams with a predetermined power splitting (PS) ratio $\gamma$: the charging beam $P_{\rm{ch}}$ and the communication beam $P_{\rm{com}}$,
\begin{equation}\label{eq:3-38}
    \begin{aligned}
    \left\{\begin{array}{l}
    P_{\rm{ch}} = \gamma P_{\rm{out}} \\
    P_{\rm{com}} = (1-\gamma) P_{\rm{out}}
    \end{array}\right..
    \end{aligned}
    \end{equation}

Then, the charging beam can be converted into electrical power to charge the receiver's battery through a PV panel. In this study, we propose the utilization of the single-diode model's equivalent circuit for the PV panel, enabling us to accurately calculate the output power for charging purposes, which reads~\cite{sera2007pv}
\begin{equation}\label{eq:3-39}
\setlength{\abovedisplayskip}{4pt}
    \setlength{\belowdisplayskip}{4pt}
    \begin{aligned}
    \left\{\begin{array}{l}
    P_{\mathrm{e}}=i_{\mathrm{pv}}^2 E_{\mathrm{pv}} \\
    i_{\mathrm{pv}}=\eta_{\mathrm{pv}} P_{\mathrm{ch}}-I_{\mathrm{o}}\left[e^{\frac{\left(v_{\mathrm{pv}}+i_{\mathrm{pv}} E_{\mathrm{s}}\right) q}{n_{\mathrm{s}} F K T}}-1\right]-\frac{v_{\mathrm{pv}}+i_{\mathrm{pv}} E_{\mathrm{s}}}{E_{\mathrm{sh}}}
    \end{array}\right.,
    \end{aligned}
    \end{equation}
where $i_{\mathrm{pv}}$ is the output current, $E_{\mathrm{pv}}$, $E_{\mathrm{s}}$, and $E_{\mathrm{sh}}$ represent the load resistance, series resistance, and parallel resistance of PV panel, respectively. $\eta_{\mathrm{pv}}$ and $v_{\mathrm{pv}}$ are the conversion responsivity and output voltage. In addition, $I_{\mathrm{o}}$, $F$, and $q$ in several express the dark saturation current, diode quality factor, and quantity of electric charge. $n_{\mathrm{s}}$ is the number of cells connected in series.

Furthermore, for communication, the signal current emitted from the APD is depicted as:
\begin{equation}\label{eq:R-10}
\setlength{\abovedisplayskip}{3pt}
    \setlength{\belowdisplayskip}{3pt}
\begin{aligned}
i_{data} = \eta_{\rm{apd}} P_{\rm{com}} = \eta_{\rm{apd}}(1-\gamma) P_{\rm{out}},
\end{aligned}
\end{equation}
with the optical-to-electrical conversion responsivity of APD $\eta_{\rm{apd}}$. Next, the spectral efficiency, which corresponds to the maximum data rate achievable from the APD, is determined using both ``Shannon theory" and the ``Theorem $7$ Eq.(26)" in~\cite{lapidoth2009capacity} for information reception, reads
\begin{equation}\label{eq:3-40}
    \begin{aligned}
    \left\{\begin{array}{l}
    C=\frac{1}{2} \log \left(1 + {\rm SNR} \right), \\
    {\rm SNR} = \frac{i_{data}^2}{2 e \pi n_{\text {total}}^2}
    \end{array}\right.,
    \end{aligned}
    \end{equation}
where ${\rm SNR}$ represents the signal-to-noise ratio for communication. $n_{\text {total }}$ is the power of an Additive White Gaussian Noise (AWGN) and can be derived by summing shot noise $n_{\text {shot }}$ and thermal noise $n_{\text {thermal }}$:
\begin{equation}\label{eq:3-40-2}
    \begin{aligned}
    n_{\text {total }}^2=n_{\text {shot }}^2+n_{\text {thermal }}^2 = \underbrace{2 q\left(\eta_{\mathrm{apd}} P_{\mathrm{com}}+I_{\mathrm{bc}}\right) B_{\mathrm{n}}}_{n_{\text {shot }}^2}+\underbrace{\frac{4 K T B_{\mathrm{n}}}{E_{\mathrm{apd}}}}_{n_{\text {thermal }}^2},
    \end{aligned}
    \end{equation}
where $I_{\mathrm{bc}}$, $B_{\mathrm{n}}$, and $E_{\mathrm{apd}}$ denote the background current, noise bandwidth, and load resistance of APD. $q$ is the electron charge. Furthermore, $K$ and $T$ in \eqref{eq:3-39} and \eqref{eq:3-40} denote Boltzmann's constant and the temperature in Kelvin.

\section{Numerical Analysis}\label{Section4}
To evaluate the performance of RIS-assisted resonant beam SWIPT system, we investigate the detection properties as well as various factors that impact the transmission quality, including the transmission distance, RIS position, and receiver placement in this section.

\begin{table*}[t]
\vspace{-1.0em}
    \renewcommand\arraystretch{0.8}
    \setlength{\abovecaptionskip}{-10pt}
    \setlength{\belowcaptionskip}{-10pt}
    \centering
     \caption{Parameters for simulation analysis}.
    \begin{tabular}{p{0.5cm}<{\centering}p{4.8cm}<{\centering}p{1.9cm}<{\centering}|p{0.5cm}<{\centering}p{4.8cm}<{\centering}p{1.8cm}<{\centering}}
    \hline
    \textbf{{\small Symbol}} & \textbf{{\small Parameter}} & \textbf{{\small Value}} & \textbf{{\small Symbol}} & \textbf{{\small Parameter}} & \textbf{{\small Value}} \\
     \hline
     \bfseries{{\small $r$}} & {{\small Radius of input/output reflectors}} & {{\small $2.5 \rm{mm}$}} & {{\small $r_{\rm{g}}$}} & {{\small Radius of reflecting mirror}} & {{\small $2.5 \rm{mm}$}} \\

    \bfseries{{\small $R_{\rm{in}}$}} & {{\small Reflectivity of input reflectors}} & {{\small $100\%$}} & {{\small $R_{\rm{out}}$}} & {{\small Reflectivity of output reflectors}} & {{\small $95\%$}} \\

    \bfseries{{\small $I_{\rm{s}}$}} & {{\small Saturation intensity}} & {\small {$1260 \rm{W/cm^2}$}} & {{\small $\eta_{\rm{excit}}$}} & {{\small Excitation efficiency}} & {{\small $72\%$}} \\

    \bfseries{{\small $\eta_{\rm{s}}$}} & {{\small Transfer efficiency of medium}} & {{\small $99\%$}} & {{\small $\lambda$}} & {{\small Wavelength of free-space beam}} & {{\small $1064 \rm{nm}$}} \\
    \hline

    \bfseries{{\small $E_{\rm{pv}}$}} & {{\small Load resistance of PV}} & {{\small $100 \Omega$}} & {{\small $E_{\rm{s}}$}} & {{\small Panel series resistance}} & {{\small $0.93 \Omega$}}\\

    \bfseries{{\small $E_{\rm{sh}}$}} & {{\small Panel parallel resistance}} & {{\small $52.6 k\Omega$}} & {{\small $\eta_{\rm{pv}}$}} & {{\small Conversion responsivity}} & {{\small $0.0161A/W$}}\\

    \bfseries{{\small $I_{\rm{o}}$}} & {{\small Dark saturation current}} & {{\small $9.89\times10^{-9} \rm{A}$}} & {{\small $F$}} & {{\small Diode quality factor}} & {{\small $1.105$}}\\

    \bfseries{{\small $n_{\rm{s}}$}} & {{\small Number of PV cell}} & {{\small $40$}} & {{\small $\eta_{\rm{apd}}$}} & {{\small Conversion efficiency}} & {{\small $0.6 \rm{A/W}$}}\\

    \bfseries{{\small $I_{\rm{c}}$}} & {{\small Background current}} & {{\small $5100 \rm{\mu A}$}} & {{\small $B_{\rm{n}}$}} & {{\small Noise bandwidth}} & {{\small $811.7 \rm{MHz}$}}\\

    \bfseries{{\small $E_{\rm{apd}}$}} & {{\small Load resistor}} & {{\small $10 \rm{k\Omega}$}} & {{\small $q$}} & {{\small Quantity of electric charge}} & {{\small $1.6\times10^{-9}$}}\\

    \bfseries{{\small $K$}} & {{\small Boltzmann constant}} & {{\small $1.38\times10^{-23}$}} & {{\small $T$}} & {{\small Kelvin Temperature}} & {{\small $300 \rm{K}$}}\\

    \hline
    \label{table2}
    \end{tabular}
    \vspace{-2.0em}
\end{table*}

The parameters of the system are displayed in Table~\ref{table2}. First, the input/output reflector and gain medium have a radius of $2.5 \rm{mm}$, and the input and output cat's eye reflectors have a focal lens of $50 \rm{mm}$. To ensure complete reflection of the resonant beam by the RIS, the size of the RIS is set to match that of the input/output reflector. Namely, the RIS size for simulation can be $5\rm{mm}\times5\rm{mm}$. The reflectivity of output reflector is $95\%$. The reflectivity of the output reflector is $95\%$. The selected gain medium is Nd:YVO$_4$, which exhibits the following specifications: a saturation intensity of $1260\rm{W/cm^2}$, a transfer coefficient of $99\%$, and an excitation efficiency of $72\%$. The wavelength of free-space beam is $1064 \rm{nm}$. Additionally, the parameters for charging and communication in equations~\eqref{eq:3-39} and~\eqref{eq:3-40} are listed in Table~\ref{table2}. The parameters for photovoltaic conversion using PV panels are derived from the measured data of a real PV cell product in~\cite{perales2016characterization}, while the parameters for the APD are obtained from experimental results in~\cite{demir2017handover}.

\begin{figure*}[t]
\vspace{-1.0em}
\centering
\subfigcapskip=-5pt
\subfigure[$d_{ex} = 0mm$]{
\begin{minipage}[t]{0.18\linewidth}
\centering
\includegraphics[width=1.5in]{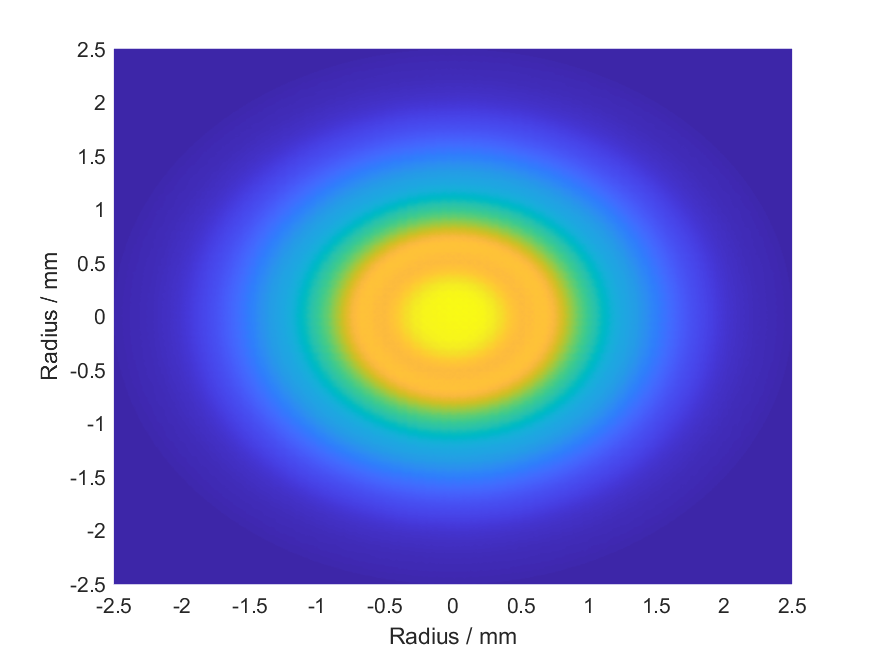}
\end{minipage}
}
\subfigure[$d_{ex} = 0.25mm$]{
\begin{minipage}[t]{0.18\linewidth}
\centering
\includegraphics[width=1.5in]{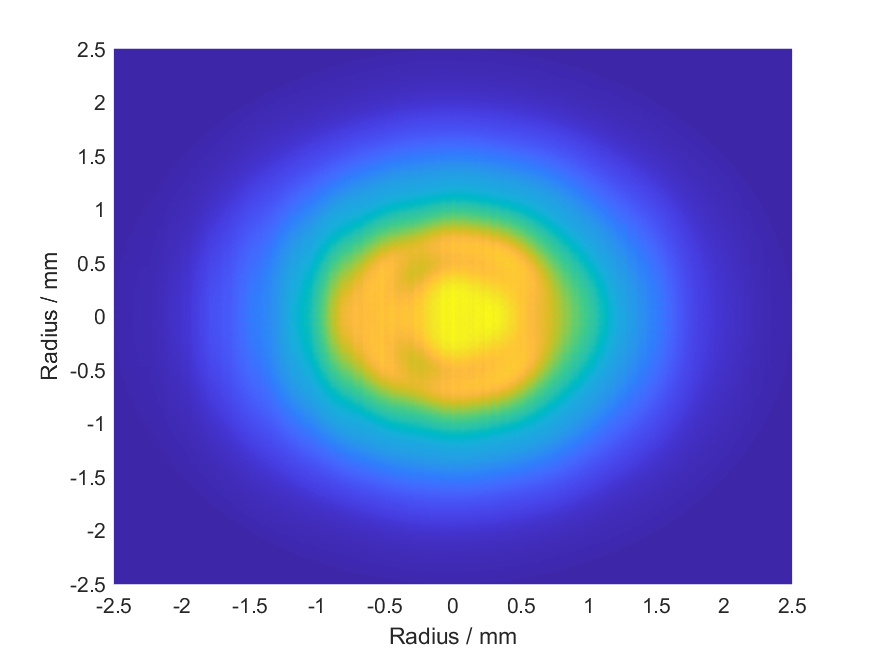}
\end{minipage}%
}%
\subfigure[$d_{ex} = 0.5mm$]{
\begin{minipage}[t]{0.18\linewidth}
\centering
\includegraphics[width=1.5in]{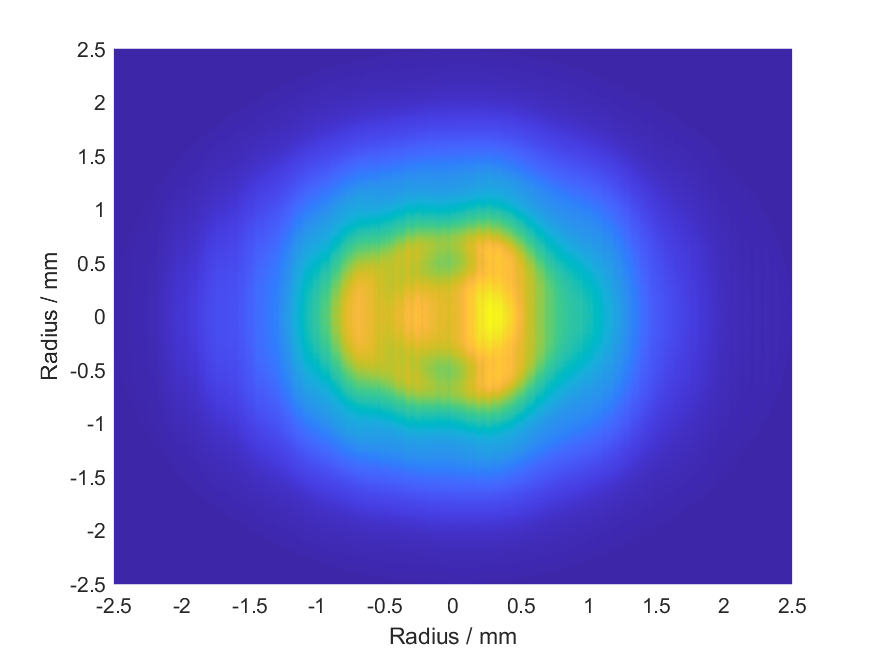}
\end{minipage}
}%
\subfigure[$d_{ex} = 0.75mm$]{
\begin{minipage}[t]{0.18\linewidth}
\centering
\includegraphics[width=1.5in]{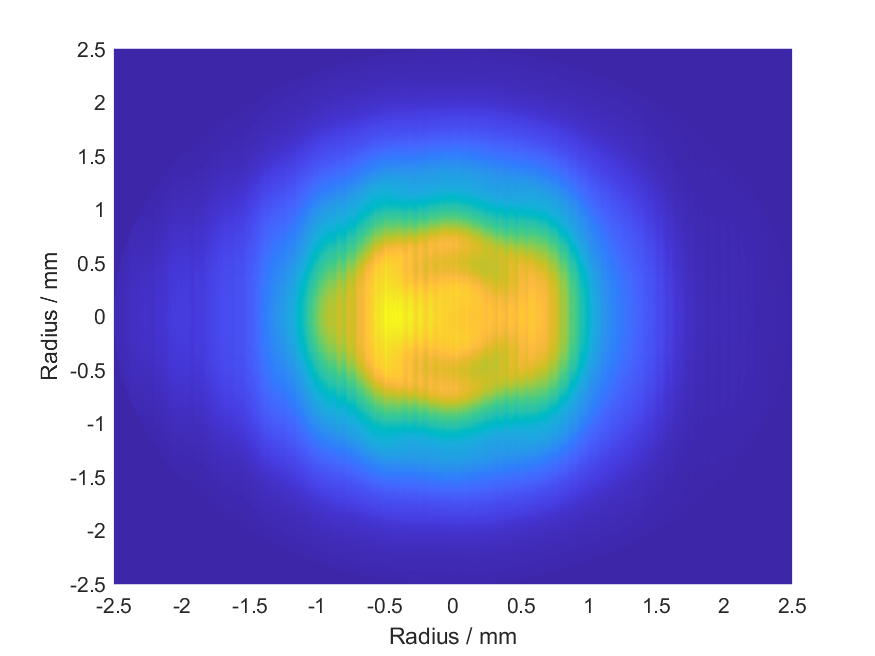}
\end{minipage}
}%
\subfigure[$d_{ex} = 1mm$]{
\begin{minipage}[t]{0.18\linewidth}
\centering
\includegraphics[width=1.5in]{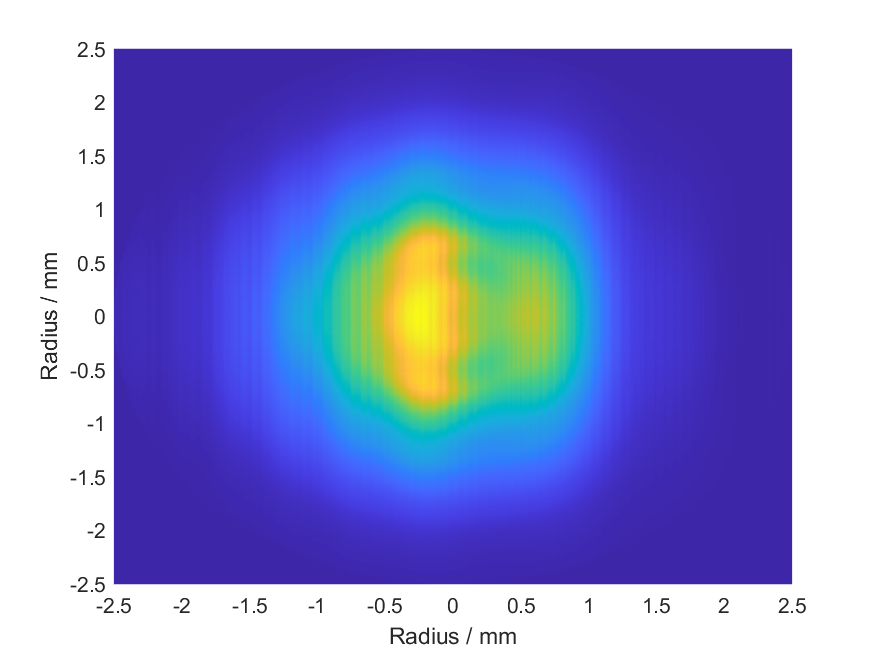}
\end{minipage}
}

\subfigure[$d_{ex} = 1.25mm$]{
\begin{minipage}[t]{0.18\linewidth}
\centering
\includegraphics[width=1.5in]{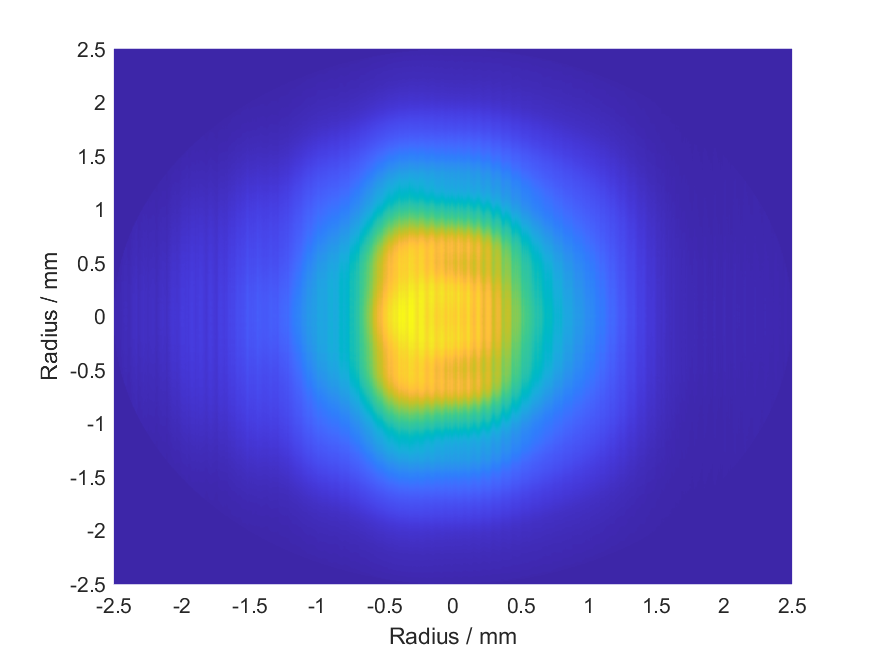}
\end{minipage}
}
\subfigure[$d_{ex} = 1.5mm$]{
\begin{minipage}[t]{0.18\linewidth}
\centering
\includegraphics[width=1.5in]{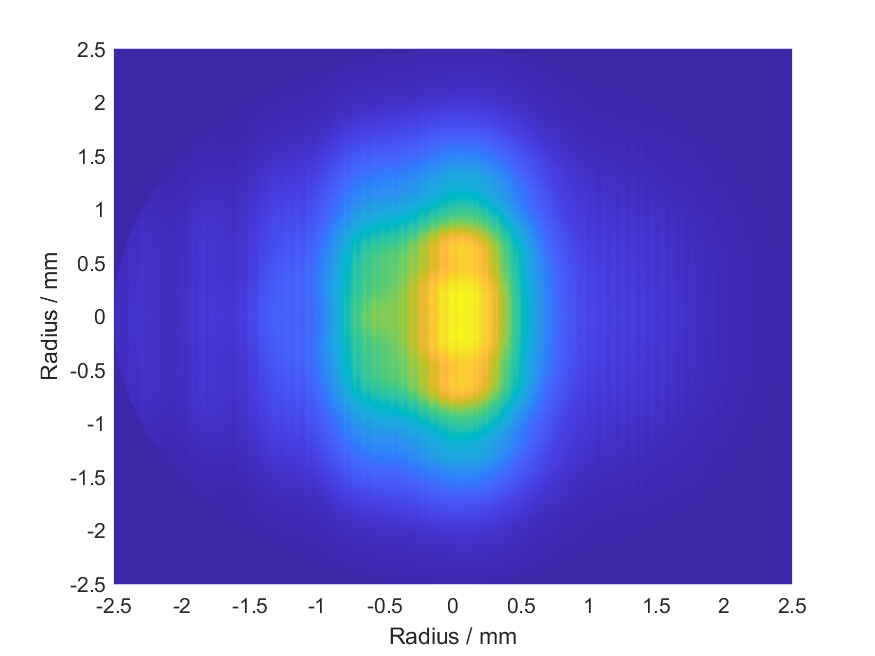}
\end{minipage}%
}%
\subfigure[$d_{ex} = 1.75mm$]{
\begin{minipage}[t]{0.18\linewidth}
\centering
\includegraphics[width=1.5in]{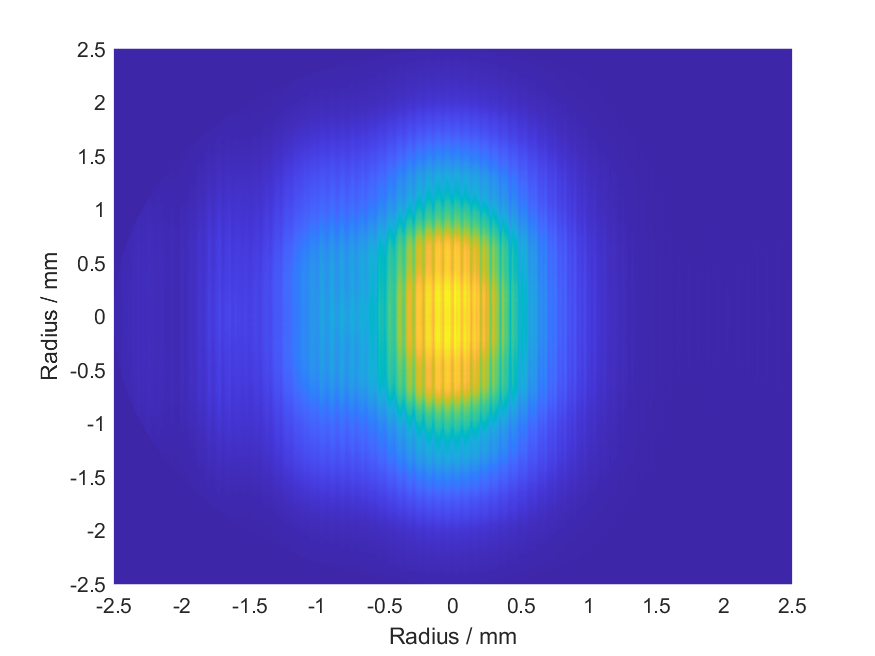}
\end{minipage}
}%
\subfigure[$d_{ex} = 2mm$]{
\begin{minipage}[t]{0.18\linewidth}
\centering
\includegraphics[width=1.5in]{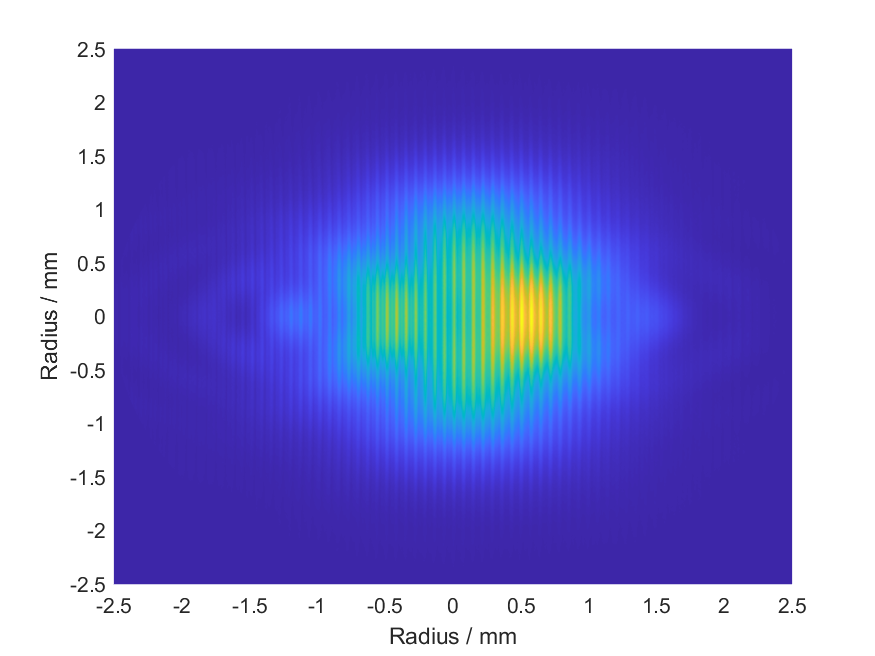}
\end{minipage}
}%
\subfigure[$d_{ex} = 2.25mm$]{
\begin{minipage}[t]{0.18\linewidth}
\centering
\includegraphics[width=1.5in]{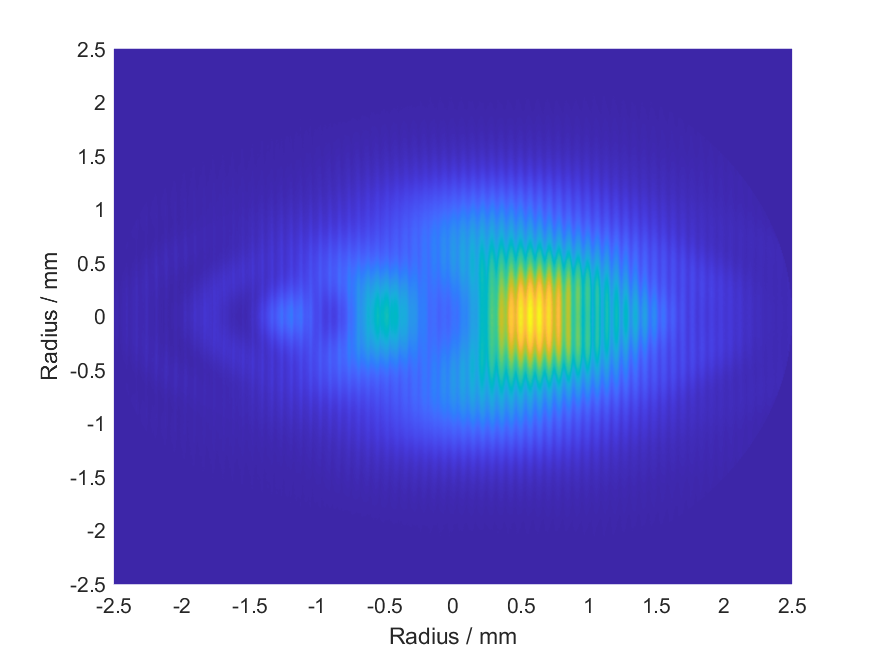}
\end{minipage}
}%
\centering
\caption{The changes of normalized field distribution on the CMOS with the invasion depth rises from $0\rm{mm}$ to $2.25\rm{mm}$.}
\label{Fig:invasiondistribution}
\vspace{-2.0em}
\end{figure*}

\vspace{-15pt}
\subsection{Obscuration Detection}\label{}
The optical field propagation will be affected if there is an invasion of an external object into the transmission channel. Considering the boundary of the reflector (i.e. a radius of $2.5 \rm{mm}$) as the invasion origin, the external object obscures the transfer apertures from the negative direction of $x$-axis. Following that, Fig.~\ref{Fig:invasiondistribution} illustrates how the increase in invasion depth alters the normalized field distribution on the CMOS in the transmitter without considering the power of transfer beam.

If the transmission channel remains unobstructed by any external objects ($d_{\rm{ex}} = 0$), the normalized field distribution follows the standard fundamental mode distribution, i.e. the Gaussian distribution as depicted in Fig.~\ref{Fig:invasiondistribution}(a). Figure~\ref{Fig:invasiondistribution}(b)-(j) demonstrates that the area of the normalized field distribution gradually reduces with increasing invasion depth, primarily due to the decrease in the effective transmission aperture size. Thus, the energy on each optical plane decreases as the distribution range of high-energy fields becomes smaller.

\begin{figure*}[t]
\vspace{-1.0em}
\setlength{\abovecaptionskip}{0pt}
\setlength{\belowcaptionskip}{-10pt}
\centering
\subfigcapskip=-10pt
\begin{minipage}[t]{0.48\linewidth}
\centering
\includegraphics[scale=0.55]{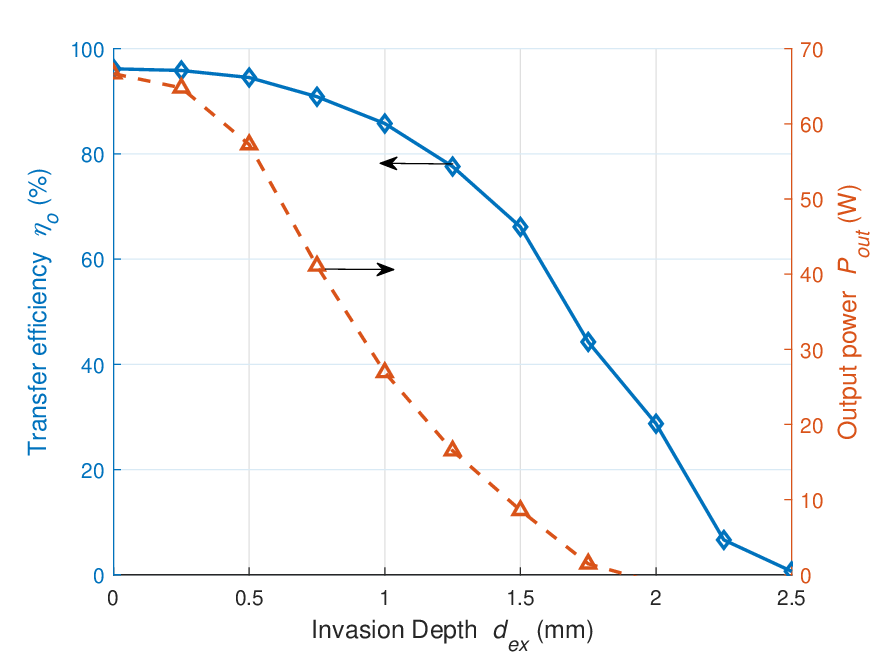}
\caption{The changes of transmission efficiency and output power with the increase in invasion depth of external object.}
\label{fig:Invasion}
\end{minipage}
\hfill
\begin{minipage}[t]{0.48\linewidth}
\centering
\includegraphics[scale=0.55]{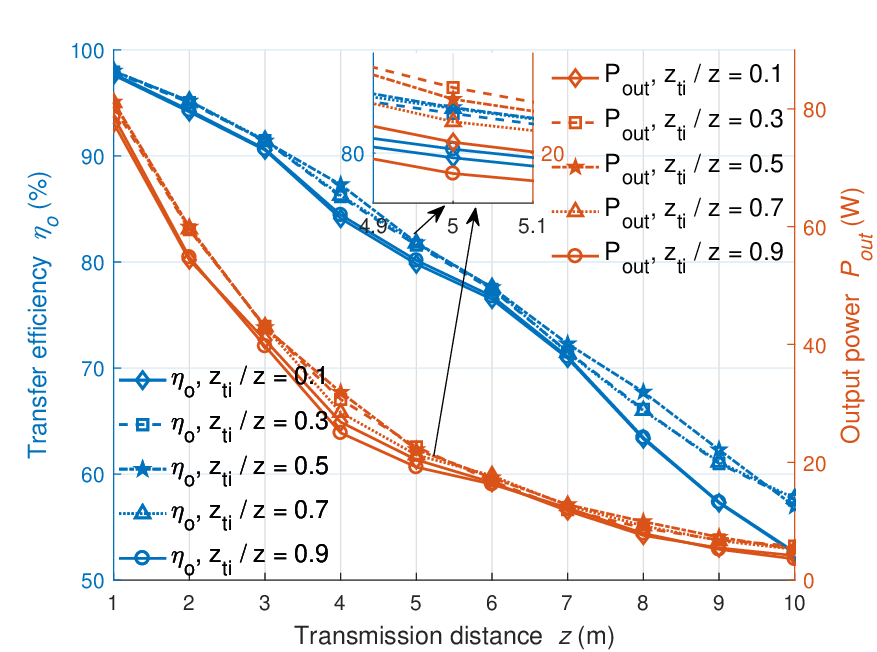}
\caption{The changes of transmission efficiency and output power as transfer distance increases (different RIS positions).}
\label{fig:dis}
\end{minipage}
\vspace{-2.0em}
\end{figure*}

Furthermore, the changes in transfer efficiency of a round-trip transmission and the output power caused by the external object invasion are illustrated in Fig.~\ref{fig:Invasion}. As the invasion depth $d_{\rm{ex}}$ increases from $0 \rm{mm}$ to $2.5 \rm{mm}$, the transfer efficiency calculated by the field distribution in the same optical plane, as given by~\eqref{eq:3-35}, progressively declines from $96\%$ to $0.6\%$. Additionally, for invasion depths below $0.5\rm{mm}$, the decrease in transfer efficiency is relatively small, as the invasion has a minimal effect on the field distribution. Moreover, the output power gradually diminishes with increasing invasion depth until it reaches zero at approximately $d_{\rm{ex}}\approx1.8\rm{mm}$. Meanwhile, if $d_{\rm{ex}}\approx 1.8\rm{mm}$, the absolute intensity distribution, which is the product of the relative field distribution and power, will also be zero.


\subsection{RIS-Assisted Transmission}\label{}
The deployment of an RIS between the transmitter and receiver presents an opportunity to enhance the efficiency of NLOS free-space optical transmission. In this subsection, we analyze the effects of various factors on system performance, including the transmission distance and the placement location of the RIS, such as the horizontal position along the $z$-axis and the vertical height relative to the $z$-axis.

\subsubsection{RIS with different horizontal positions}

\

\begin{figure*}[t]
\vspace{-1.0cm}
\setlength{\abovecaptionskip}{0cm}
\centering
\subfigcapskip=-5pt
\subfigure[Transfer efficiency and output power]{
\centering
\includegraphics[width=0.45\textwidth]{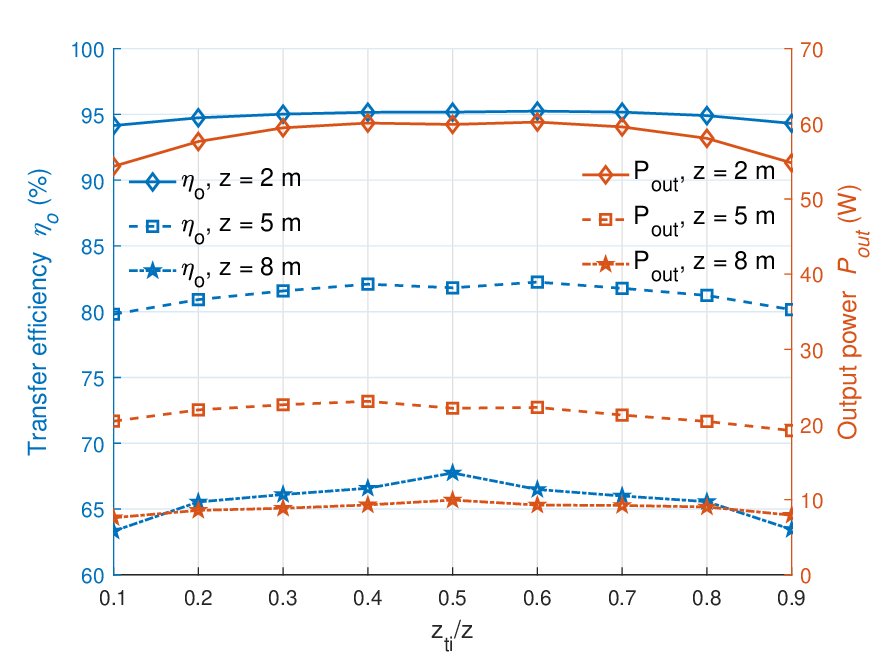}
\label{fig:position}
}
\subfigure[Received power and SNR]{
\centering
\includegraphics[width=0.45\textwidth]{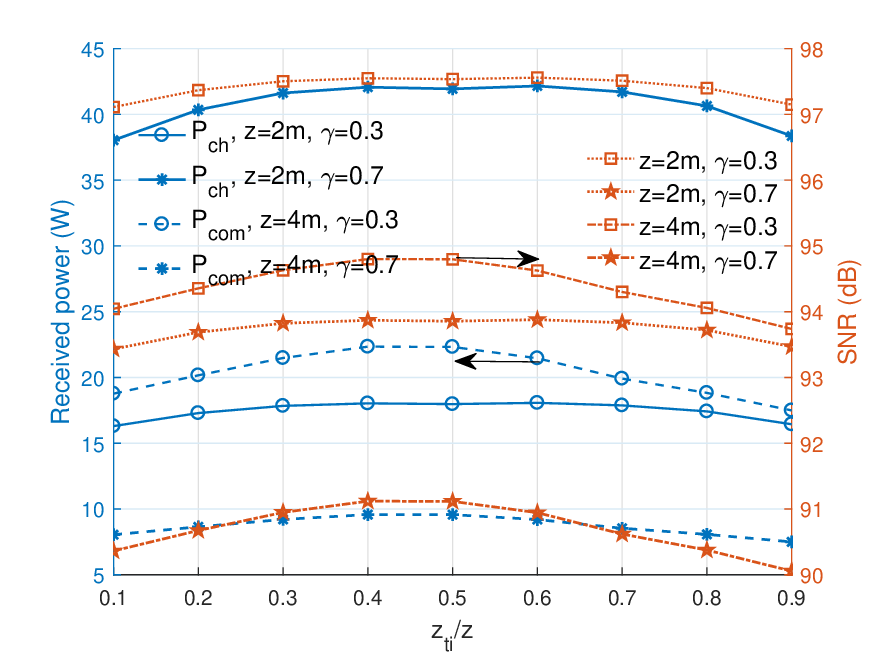}
\label{fig:hoSNR}
}
\\ \vspace{-8pt}
\subfigure[SWIPT performance]{
\centering
\includegraphics[width=0.45\textwidth]{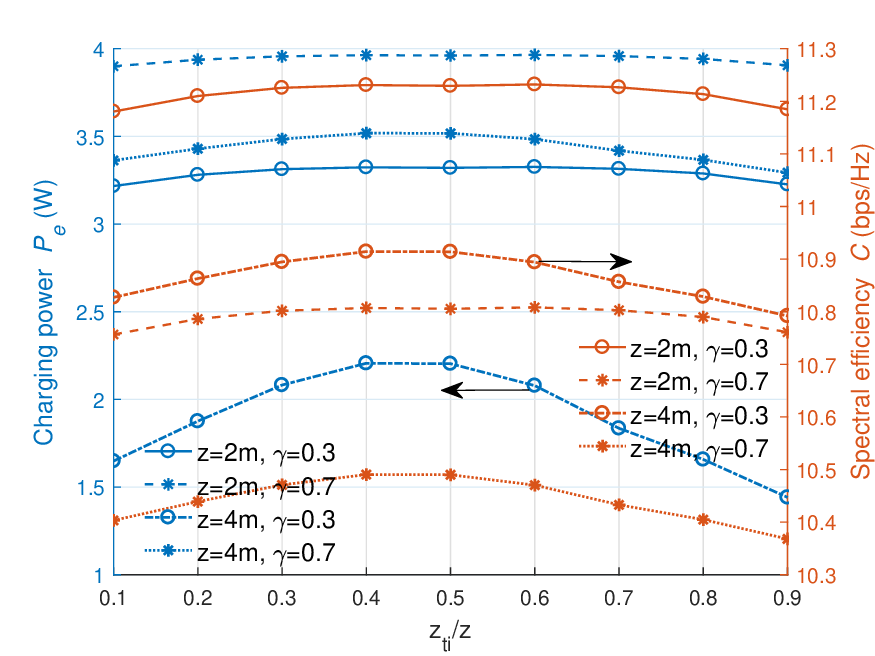}
\label{fig:swiptposition}
}
\subfigure[Tradeoff of SWIPT]{
\centering
\includegraphics[width=0.45\textwidth]{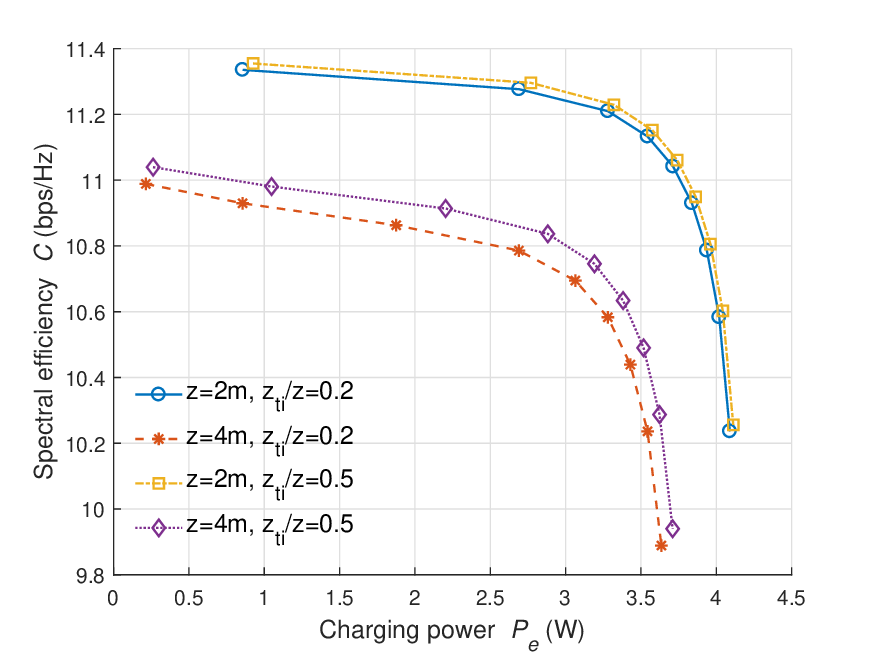}
\label{fig:horitradeoff}
}
\centering
\caption{The changes of transfer performance with various RIS horizontal positions, transfer distances, and PS ratios.}
\label{Fig:horizontal}
\vspace{-1.0cm}
\end{figure*}

As depicted in \eqref{eq:3-12}, \eqref{eq:3-14}, and \eqref{eq:3-17}, we can see the distance $z_{\rm{ti}}$ between the transmitter and RIS, along with the vertical height $d_{\rm{iz}}$ from the RIS to the $z$-axis, primarily impact the transfer distance and phase angle. The variations in transfer efficiency and output power in relation to the changes in transfer distance $z$ and $z_{\rm{ti}}$ are depicted in Figs.~\ref{fig:dis} and~\ref{fig:position}, assuming a fixed vertical height of $0.5\rm{m}$ from the RIS to the $z$-axis.

Firstly, the diffraction loss increases with the growth of transmission distance. Thus, the transfer efficiency in a round-trip transmission $\eta_{\rm{o}}$ declines from $97.63\%$ to $52.54\%$ slowly as the transmission distance $z$ rises from $1 \rm{m}$ to $10 \rm{m}$ in Fig.~\ref{fig:dis}. As a result, the output power, which is proportional to the transfer efficiency and calculated in (36), also decreases gradually. For instance, when $z = 2\rm{m}$ and $4\rm{m}$ with a ratio of the distance between the transmitter and RIS along the $z$-axis to the transmission distance set at $0.5$, the transfer efficiency $\eta_{\rm{o}}$ is $95.18\%$ and $87.30\%$, and the corresponding output powers $P_{\rm{out}}$ are $59.92\rm{W}$ and $22.15\rm{W}$, respectively. Moreover, the maximum output beam power is about $80 \rm{W}$ at $z = 1 \rm{m}$ and $z_{\rm{ti}}/z = 0.5$. Additionally, if the RIS is positioned at the center of the resonant cavity ($z_{\rm{ti}}/z = 0.5$), both the transfer efficiency and output power surpass those obtained at other positions for the same transmission distance. However, due to field distribution calculation errors, the difference in transmission efficiency and power between the same RIS locations at various distances is different. Finally, the maximal energy efficiency is about $55\%$ ($80/(200*72\%)\approx 55\%$).

The impact of horizontal RIS placement is illustrated in Fig.~\ref{fig:position}, where $z_{\rm{ti}}/z$ denotes the ratio of the distance between the transmitter and RIS along the $z$-axis to the transmission distance. For example, if $z = 5 \rm{m}$ and $z_{\rm{ti}}/z = 0.3$, the distance between the transmitter and RIS along the $z$-axis is $1.5 \rm{m}$. It can be observed that the horizontal position of the RIS has a relatively minor influence on the transfer efficiency and output power. However, if the RIS is positioned closer to the center of the $z$-axis, the transfer efficiency improves slightly, and the output power is relatively higher. For example, with a transfer distance of $8\rm{m}$ and $z_{\rm{ti}}/z$ values of $0.2$, $0.5$, and $0.8$, the transfer efficiencies $\eta_{\rm{o}}$ are $65.54\%$, $67.74\%$, and $65.56\%$, respectively. The corresponding output powers are $8.57\rm{W}$, $9.94\rm{W}$, and $9.02\rm{W}$.

Then, the effect of horizontal RIS placement can be seen in Fig.~\ref{fig:position}, where $z_{\rm{ti}}/z$ represents the ratio of the distance between transmitter and RIS in $z$-axis $z_{\rm{ti}}$ and transfer distance $z$. For instance, the distance between transmitter and RIS in $z$-axis is $1.5 \rm{m}$ if $z=5 \rm{m}$ and $z_{\rm{ti}}/z = 0.3$. It can be seen that the influence of the horizontal position of RIS on transfer efficiency and output power is relatively small. However, if RIS is positioned near the center of $z$-axis, the transfer efficiency is slightly more effective, and the output power is relatively larger. For example, the transfer efficiency $\eta_{\rm{o}}$ is $65.54\%$, $67.74\%$, $65.56\%$, and the output power is $8.57 \rm{W}$, $9.94 \rm{W}$, $9.02 \rm{W}$ if $z_{\rm{ti}}/z = 0.2$, $0.5$ and $0.8$ with $8 \rm{m}$ transfer distance. This is because RIS dividing the long transmission distance into two transmission segments, thereby partially compensating for losses incurred during long-distance transmission.

%
%
Fig.~\ref{fig:hoSNR} illustrates the variations in received power for charging $P_{\rm{ch}}$ and communication $P_{\rm{com}}$ with different horizontal positions, transfer distances, and PS ratios. Increasing the power splitting factor $\gamma$ leads to higher $P_{\rm{ch}}$ but lower $P_{\rm{com}}$. Meanwhile, the SNR decreases as $\gamma$ increases for the same transmission distance and horizontal position. Similar to the trend observed in output power, SNR decreases as the transmission distance increases under the same PS ratio and horizontal position. Additionally, both received power and SNR reach a peak value and then decrease as $z_{\rm{ti}}/z$ increases. The maximum value can be reached at $z_{\rm{ti}}/z=0.5$. The maximum SNR is about $97.54{\rm{dB}}$ at $z=2m$, $\gamma=0.3$, and $z_{\rm{ti}}/z=0.5$.

The effect of the horizontal position of RIS on the SWIPT performance is presented in Fig.~\ref{fig:swiptposition}, in which the transfer distances considered are $2 \rm{m}$ and $4 \rm{m}$, with a fixed vertical height of the RIS at $0.5 \rm{m}$. Similar to the trends observed in transfer efficiency and output power, both the charging power and spectral efficiency decrease as the transfer distance $z$ increases. Moreover, if the RIS is positioned closer to the midpoint of the $z$-axis, the charging power and spectral efficiency are higher. That is, $P_{\rm{e}}$ and $C$ at $z_{\rm{ti}}/z=0.5$ are greater than that in $z_{\rm{ti}}/z=0.1$ and $0.9$. For example, $P_{\rm{e}}=2.20 \rm{W}$, $C=10.91 \rm{bps/Hz}$ at $z_{\rm{ti}}/z=0.5$ while they are $1.65 \rm{W}$, $10.83 \rm{bps/Hz}$ and $1.44 \rm{W}$, $10.79 \rm{bps/Hz}$ at $z_{\rm{ti}}/z=0.1$ and $0.9$ with $z=4 \rm{m}$ and $\gamma=0.3$. Besides, based on~\eqref{eq:3-38}, the charging power increases as the PS ratio $\gamma$ rises, while the spectral efficiency decreases.

Finally, the relationship between the charging power and spectral efficiency with different RIS horizontal positions, transfer distances, and PS ratios is depicted in Fig.~\ref{fig:horitradeoff}. The results indicate an inverse correlation between charging power and spectral efficiency: as the charging power increases, the spectral efficiency decreases. Additionally, the charging power and spectral efficiency are lower at longer transmission distances if the horizontal location of RIS is fixed. If $z_{\rm{ti}}/z=0.2$, $P_{\rm{e}}$ and $C$ at $2\rm{m}$ are higher compared to those at $4\rm{m}$. Additionally, placing the RIS at the midpoint of the $z$-axis results in a more efficient system. For example, the charging power and spectral efficiency are higher at $z_{\rm{ti}}/z=0.5$ compared to $z_{\rm{ti}}/z=0.2$ under the same transfer distance.

\begin{figure*}[t]
\vspace{-1.0em}
\setlength{\abovecaptionskip}{0pt}
\setlength{\belowcaptionskip}{-10pt}
\centering
\subfigcapskip=-5pt
\subfigure[Transfer efficiency and output power]{
\centering
\includegraphics[width=0.45\textwidth]{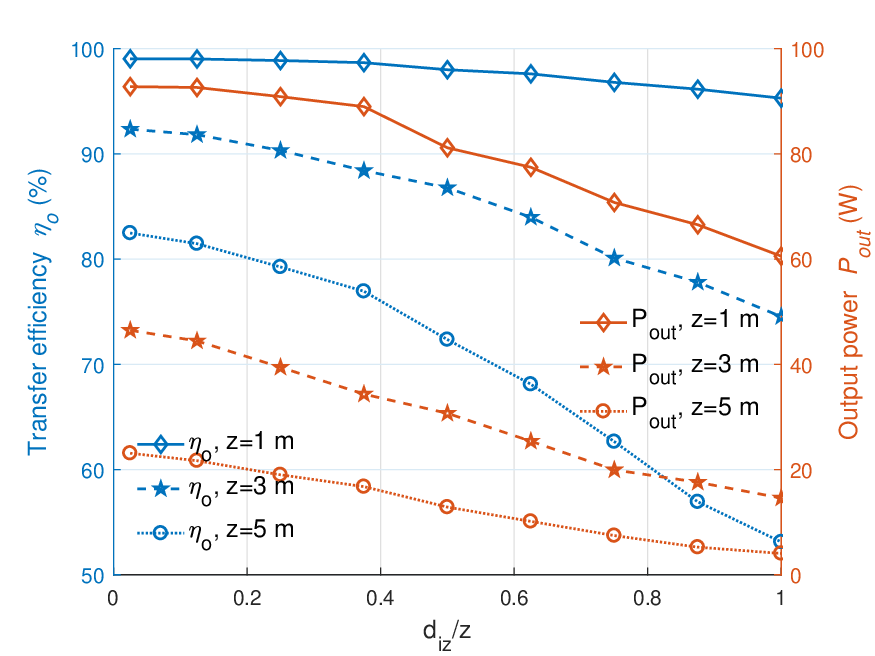}
\label{fig:height}
}
\subfigure[Received power and SNR]{
\centering
\includegraphics[width=0.45\textwidth]{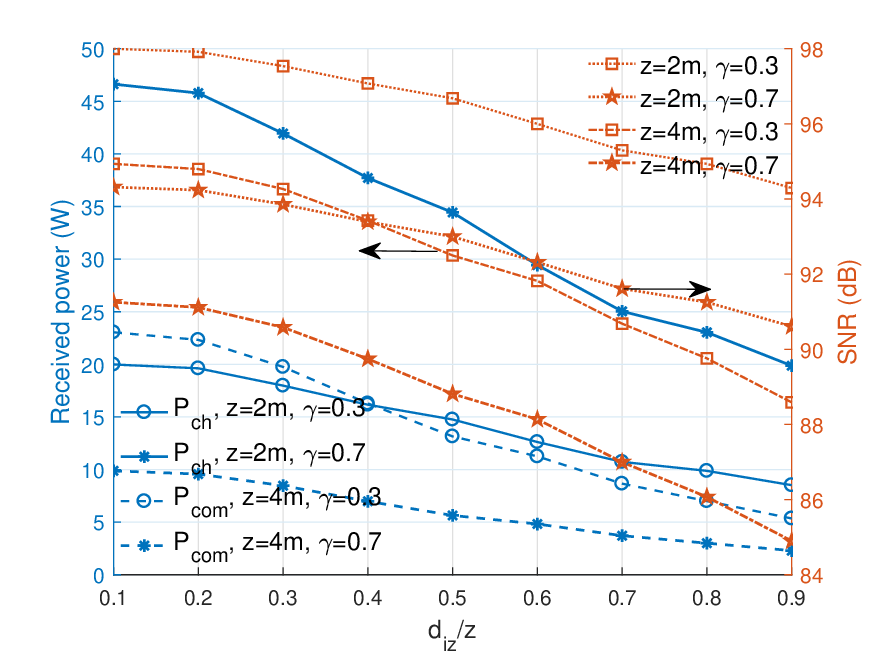}
\label{fig:heightSNR}
}
\\ \vspace{-8pt}
\subfigure[SWIPT performance]{
\centering
\includegraphics[width=0.45\textwidth]{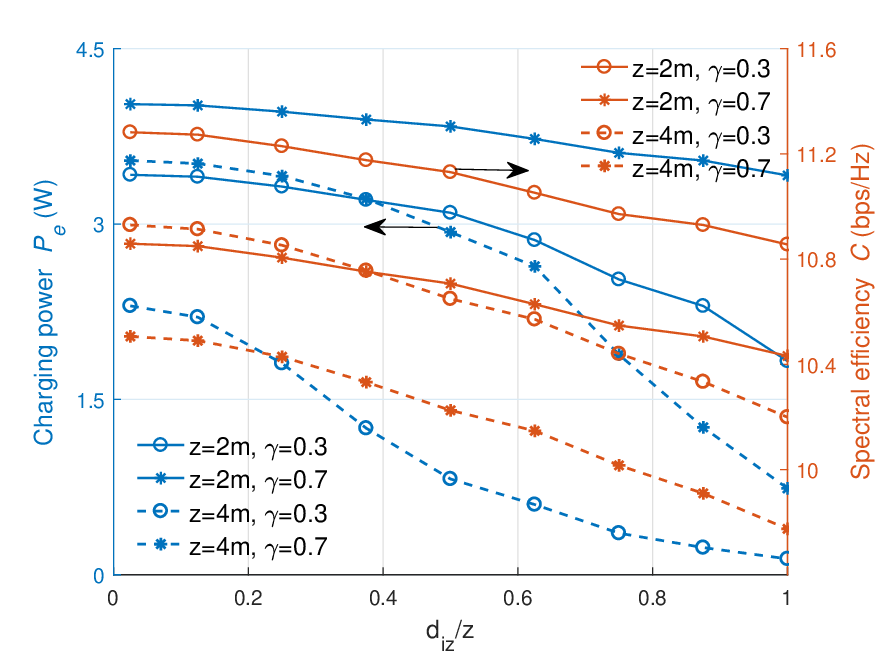}
\label{fig:swiptheight}
}
\subfigure[Tradeoff of SWIPT]{
\centering
\includegraphics[width=0.45\textwidth]{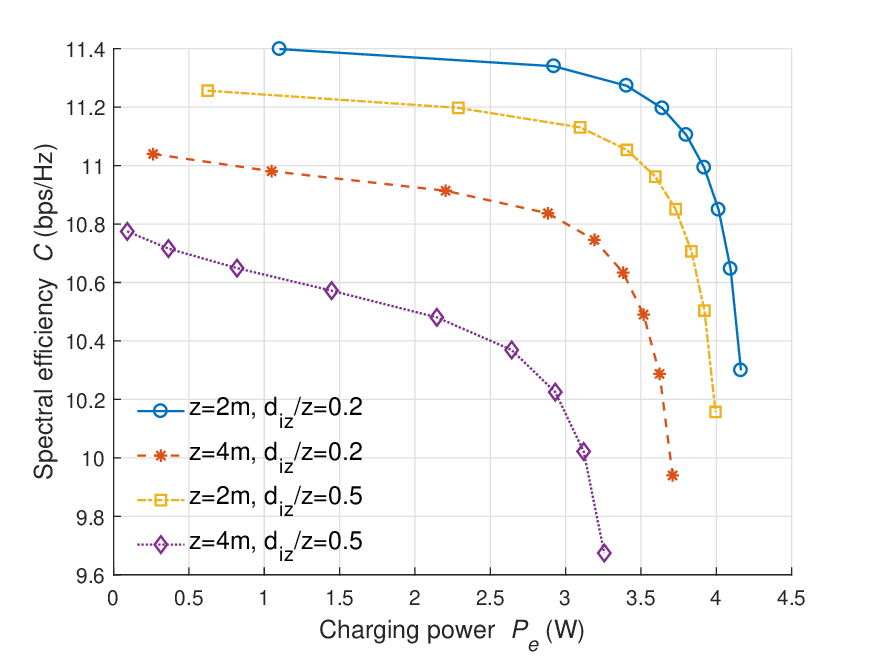}
\label{fig:heighttradeoff}
}
\centering
\caption{The changes of transfer performance with different RIS vertical heights, transfer distances, and PS ratios.}
\label{Fig:Heightperformance}
\vspace{-2.0em}
\end{figure*}

\subsubsection{RIS with different vertical heights}

\

The effect of RIS vertical height to $z$-axis $d_{\rm{iz}}$ is depicted in Fig.~\ref{fig:height}, where RIS is positioned at the midpoint of $z$-axis, i.e. $z_{\rm{ti}}/z = 0.5$. In addition, the horizontal coordinate in Fig.~\ref{fig:height} is the ratio of the vertical height $d_{\rm{iz}}$ to the transfer distance $z$. For instance, if $d_{\rm{ti}}/z = 0.2$ and the transfer distance is $3\rm{m}$, the height is $0.6\rm{m}$. The transmission loss increases with a larger height due to the increased channel distance for the free-space beam based on \eqref{eq:3-12}, \eqref{eq:3-14}, and \eqref{eq:3-17}. Thus, the transfer efficiency decreases as the height increases. Fig.~\ref{fig:height} exhibits two notable patterns: i) the decline in efficiency is relatively small when the height is small ($d_{\rm{iz}}/z < 0.4$); ii) the decrease in efficiency becomes more pronounced as the height increases, particularly for longer transmission distances. For example, $\eta_{\rm{o}}$ decreases from $76.94\%$ to $53.11\%$ with $d_{\rm{iz}}/z=0.4$ and $1$ if the transfer distance $z=5\rm{m}$, whereas $\eta_{\rm{o}}$ is $88.41\%$ and $74.54\%$ at the same height and $3 \rm{m}$ distance. Meanwhile, the output power follows a decreasing trend in line with transfer efficiency.

The trends observed in Fig.~\ref{fig:heightSNR} for received power and SNR concerning transmission distance and PS ratio are consistent with those shown in Fig.~\ref{fig:hoSNR}. Additionally, the received power for charging $P_{\rm{ch}}$ and communication $P_{\rm{com}}$ decreases as the ratio of vertical height of the RIS to the transfer distance $d_{\rm{iz}}/z$ increases. Subsequently, the SNR declines as the vertical height of RIS increases, regardless of $z$ and $\gamma$. The maximum SNR is $98{\rm{dB}}$ at $z=2\rm{m}$, $\gamma=0.3$ and $d_{\rm{iz}}/z=0.1$ in Fig.~\ref{fig:heightSNR}.

The changes in charging power and spectral efficiency with the vertical height are depicted in Fig.~\ref{fig:swiptheight}. Similar to the observed output power trend in Fig.~\ref{fig:height}, the charging power and spectral efficiency also decrease with the growth of $d_{\rm{iz}}/z$. Furthermore, increasing the transfer distance leads to a decrease in the charging power and spectral efficiency, whereas a higher PS ratio $\gamma$ results in an increase in the charging power and a decrease in the spectral efficiency. For instance, $P_{\rm{e}}=3.32 \rm{W}$, $2.86 \rm{W}$, and $C=11.23 \rm{bps/Hz}$, $11.05 \rm{bps/Hz}$ if $d_{\rm{iz}}/z=0.3$ and $0.6$ with $2 \rm{m}$ transfer distance and $\gamma=0.3$, whereas they are $1.80 \rm{W}$, $0.60 \rm{W}$, $10.85 \rm{bps/Hz}$, and $10.57 \rm{bps/Hz}$ with $4 \rm{m}$ distance. Besides, if $\gamma=0.7$ and $z=4 \rm{m}$, the charging power and spectral efficiency $3.41 \rm{W}$, $2.64 \rm{W}$, $10.43 \rm{bps/Hz}$, and $10.15 \rm{bps/Hz}$ respectively with the same parameters. The maximal charging power and spectral efficiency can reach $4.03 \rm{W}$ and $11.28 \rm{bps/Hz}$ respectively with $2 \rm{m}$ transfer distance at $\gamma=0.7$ if the horizontal position $z_{\rm{ti}}=1 \rm{m}$ and vertical height $d_{iz} = 0.05 \rm{m}$.

Similarly, as seen in Fig.~\ref{fig:heighttradeoff}, the spectral efficiency declines as spectral efficiency rises. That is, it is required to make a tradeoff between the two parameters taking into account the PS ratio $\gamma$ to optimize the charging and communication performance. Additionally, the performance degradation with increasing transfer distance is consistent with the findings in Fig.~\ref{fig:horitradeoff}. Furthermore, as the vertical height increases, the charging power and spectral efficiency decrease under the same transfer distance. For instance, $P_{\rm{e}}$ and $C$ are higher at $d_{\rm{iz}}/z=0.2$ than that at $d_{\rm{iz}}/z=0.5$, regardless of $z=2 \rm{m}$ and $4 \rm{m}$.

\begin{figure*}[t]
\vspace{-1.0em}
\centering
\subfigcapskip=-5pt
\setlength{\abovecaptionskip}{0pt}
\setlength{\belowcaptionskip}{-10pt}
\subfigure[Transfer efficiency and output power]{
\begin{minipage}[t]{0.45\linewidth}
\centering
\includegraphics[scale=0.45]{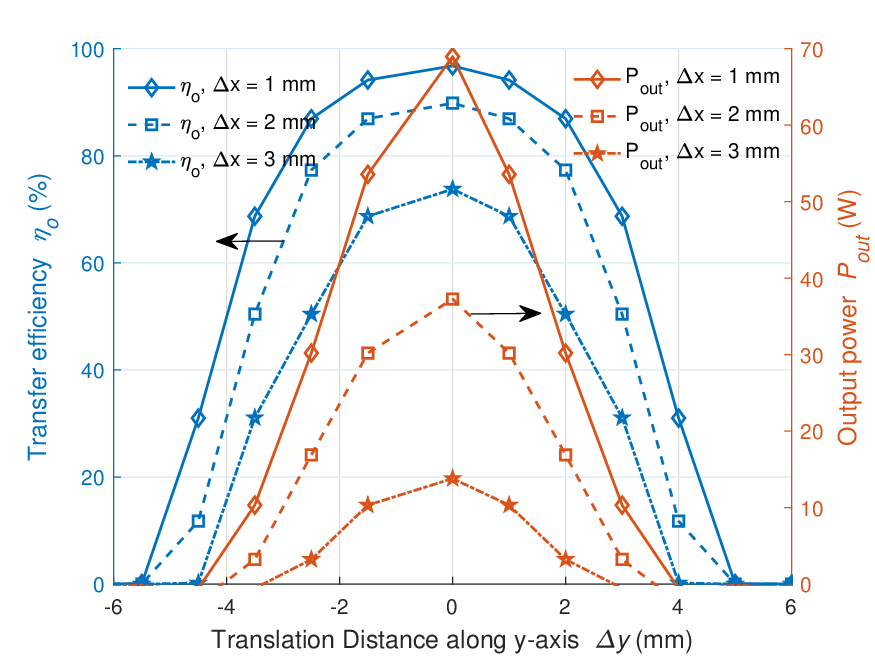}
\end{minipage}
}
\subfigure[Charging power and spectral efficiency]{
\begin{minipage}[t]{0.45\linewidth}
\centering
\includegraphics[scale=0.45]{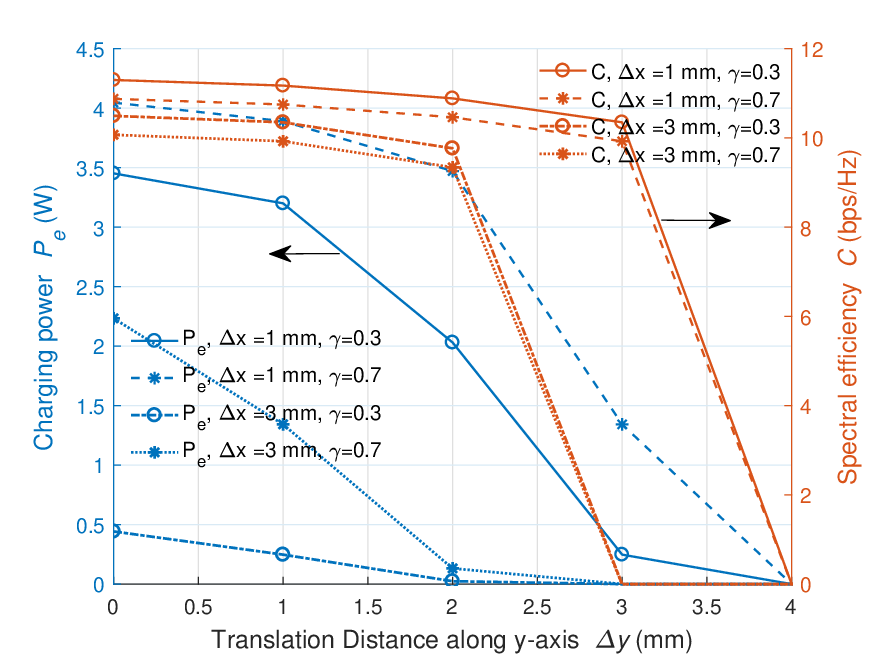}
\end{minipage}
}
\centering
\caption{The performance of RIS-assisted resonant beam S-SWIPT system with the receiver translation along the horizontal direction (i.e. the $xoy$ plane).}
\label{Fig:translation}
\vspace{-2.0em}
\end{figure*}

\vspace{-1.0em}
\subsection{RIS-Assisted Performance with Receiver Movement}\label{}
The translation or rotation of receiver causes changes in the optical field distribution on the reflectors and RIS, which in turn affect the transfer efficiency and output power. The diversity of transfer efficiency $\eta_{\rm{o}}$ and output power $P_{\rm{out}}$ with $1 \rm{m}$ transfer distance is illustrated in Fig.~\ref{Fig:translation}(a), where $z_{ti}/z=0.5$ and $d_{iz}/z=0.2$. First, if the receiver translates an equal distance along the $xoy$ plane in both the forward and reverse directions, the transfer efficiency and output power are the same. Then, if the receiver is translated along the $y$-axis, $\eta_{\rm{o}}$ and $P_{\rm{out}}$ drops from the peak value down to zero with the increase of translation distance $|\Delta y|$, regardless of whether the receiver is moved along the $x$-axis. The effective translation distance (i.e. $P_{\rm{out}}>0$) is $\pm4 \rm{mm}$ in Fig.~\ref{Fig:translation}(a). Similarly, $\eta_{\rm{o}}$ and $P_{\rm{out}}$ decreases with the extension of $\Delta x$ with the certain $\Delta y$. For example, $\eta_{\rm{o}}=86.93\%$, $77.33\%$, $50.43\%$ $P_{\rm{out}}=30.21 \rm{W}$, $16.89 \rm{W}$, and $3.23 \rm{W}$ with $\Delta x = 1 \rm{mm}$, $2 \rm{mm}$ and $3 \rm{mm}$ under $\Delta y = 2 \rm{mm}$.

Accordingly, as the translation distance along $y$-axis increases from zero to $4 \rm{mm}$, the charging power and spectral efficiency decline to zero as shown in Fig~\ref{Fig:translation}(b). The charging power and spectral efficiency with $\Delta x=2 \rm{mm}$ is greater than that at $\Delta x=3 \rm{mm}$. For instance, if $\gamma=0.3$, $P_{\rm{e}}=3.20 \rm{W}$, $0.25\rm{W}$ and $C=11.17 \rm{bps/Hz}$, $10.35 \rm{bps/Hz}$ with $\Delta x = 1 \rm{mm}$ and $\Delta x = 3 \rm{mm}$ at $\Delta y = 1 \rm{mm}$, while $P_{\rm{e}}=2.03 \rm{W}$, $0.02 \rm{W}$ and $C=10.89\rm{bps/Hz}$, $9.77 \rm{bps/Hz}$ with $\Delta x = 1 \rm{mm}$ and $\Delta x = 3 \rm{mm}$ at $\Delta y = 2 \rm{mm}$. Similar to the SWIPT performance analysis, the charging power and spectral efficiency for moving receivers are negatively correlated with respect to $\gamma$.

\begin{figure*}[t]
\vspace{-1.0em}
\centering
\subfigcapskip=-5pt
\setlength{\abovecaptionskip}{0pt}
\setlength{\belowcaptionskip}{-10pt}
\subfigure[Transfer efficiency and output power]{
\begin{minipage}[t]{0.45\linewidth}
\centering
\includegraphics[scale=0.45]{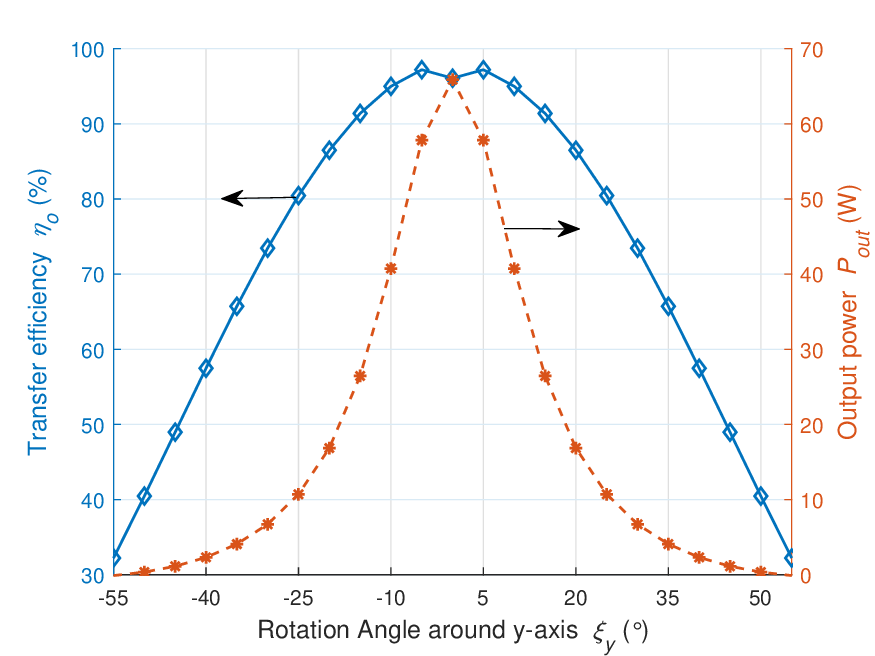}
\end{minipage}
}
\subfigure[Charging power and spectral efficiency]{
\begin{minipage}[t]{0.45\linewidth}
\centering
\includegraphics[scale=0.45]{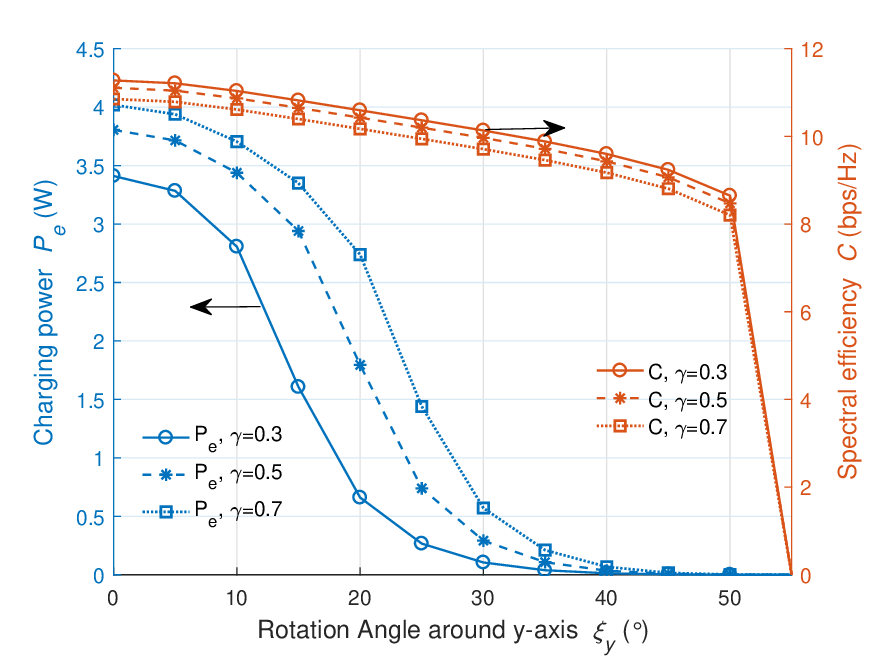}
\end{minipage}%
}
\centering
\caption{The performance of RIS-assisted resonant beam S-SWIPT system with the receiver rotation around $y$-axis.}
\label{Fig:rotation}
\vspace{-2.0em}
\end{figure*}

Figure~\ref{Fig:rotation} depicts the changes in transfer efficiency, output power, charging power, and spectral efficiency as the receiver rotates around $y$-axis. Firstly, the transfer efficiency $\eta_{\rm{o}}$ and output power $P_{\rm{out}}$ decrease gradually with the growth of the absolute value of the rotation angle $\xi_y$ in Fig.~\ref{Fig:rotation}(a). Both the values fall symmetrically if the forward and reverse rotation angles are the same. $\eta_{\rm{o}}=94.99\%$, $86.50\%$ and $P_{\rm{out}}=40.75 \rm{W}$, $16.86 \rm{W}$ with $\xi_y = 10^\circ$ and $20^\circ$. Additionally, the output power decreases to zero at $\xi_y \approx 50^\circ$. That is, the effective rotation angle around $y$-axis of receivers is $\pm 50^\circ$. In Fig.~\ref{Fig:rotation}(b), the charging power and spectral efficiency decrease from the peak power to zero as the escalation of rotation angle $\xi_y$. In addition, if the PS ratio $\gamma$ increases, the charging power rises while the spectral efficiency declines. $P_{\rm{e}}=0.66 \rm{W}$, $1.80 \rm{W}$, $2.74 \rm{W}$ and $C=10.60 \rm{bps/Hz}$, $10.43 \rm{bps/Hz}$, $10.17 \rm{bps/Hz}$ with $\gamma=0.3$, $0.5$, and $0.7$ at $\xi_y=20^\circ$.


Overall, the detection of external object intrusion can be rapidly achieved by monitoring changes in the optical field distribution. Furthermore, RIS plays a vital role in enabling efficient NLOS mobile optical transmission.

\section{Conclusions And Discussions}\label{Section5}
\subsection{Conclusions}
To address the limitation of requiring LoS links between the transmitter and the receiver, we propose a RIS-assisted resonant beam SWIPT system. This system enables sensitive detection of obscuration in the transfer channel and facilitates high-power NLOS transmission with the assistance of RIS. By applying electromagnetic wave propagation theory, we can derive the optical field distributions in each plane for the transmitter-RIS (T-RIS) and RIS-receiver (RIS-R) channels. Through numerical analysis, we demonstrate that the presence of external objects results in a decrease in transfer efficiency and output power, which can be detected sensitively by monitoring changes in the optical field distribution. The effects of RIS on the transmission performance depend on its horizontal positions and vertical heights. If the RIS is positioned closer to the midpoint of the transceiver axis and its vertical height is approximately equal to the radius of the reflector, the output power is almost maximum. Furthermore, the proposed system facilitates mobile energy transfer within $8 \rm{mm}$ translation distance and $100^\circ$ rotation angle, and can achieve above $4 \rm{W}$ charging power and $11 \rm{bps/Hz}$ channel capacity.
\vspace{-15pt}

\subsection{Discussions}\label{}
Moreover, there remain certain unresolved issues pertaining to RIS-assisted resonant beam transmission that warrant further investigation. These issues include the following:
\begin{itemize}
  \item Communication Performance
\end{itemize}

\begin{figure}[!t]
\vspace{-1.0em}
\setlength{\abovecaptionskip}{-5pt}
\setlength{\belowcaptionskip}{-10pt}
    \centering
    \includegraphics[scale=0.6]{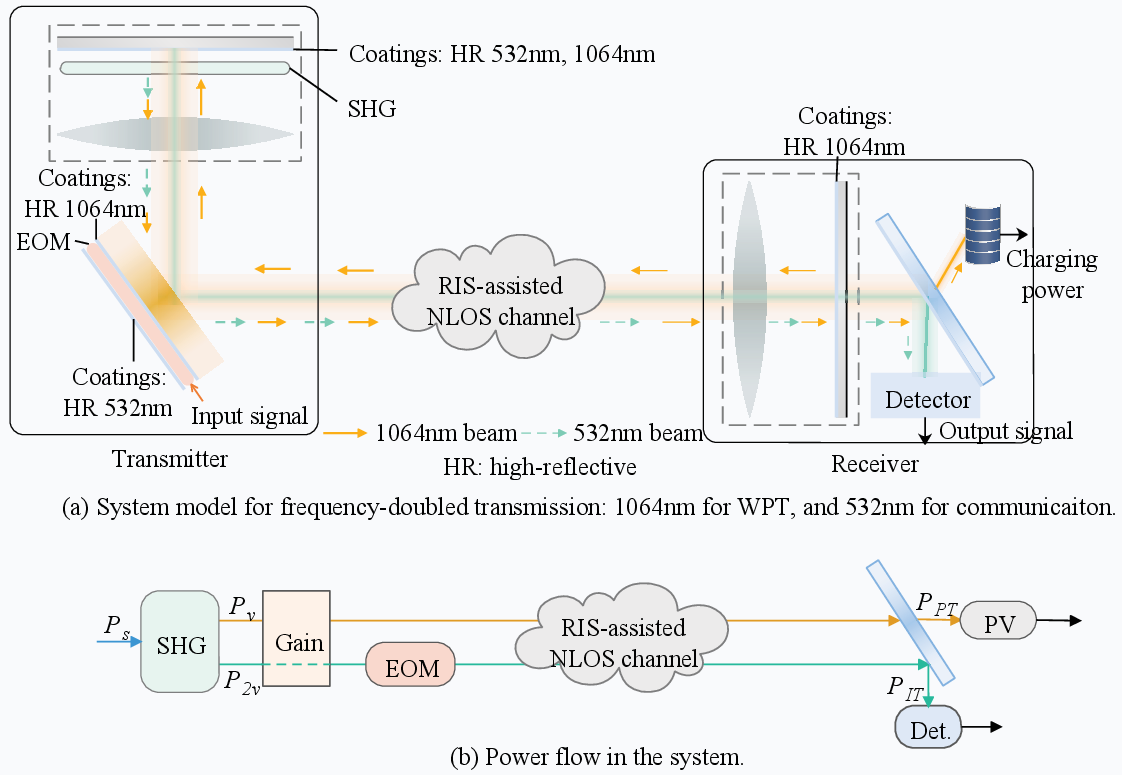}
    \caption{The example structure of RIS-assisted resonant beam transmission system using the frequency-doubling method.}
    \label{fig:doublefre}
    \vspace{-2.0em}
\end{figure}

To provide a high-power and high-capacity mobile transmission channel for IoT devices, we adopt the second harmonic generation (SHG) and electro-optic modulator (EOM) modules in transmitter of the RIS-assisted resonant beam system~\cite{xiong2021mobile}. As illustrated in Fig.\ref{fig:doublefre}(a), the resonant beam undergoes frequency doubling via SHG to generate a communication beam at frequency-doubled, which is then modulated by the EOM. The design of coatings with varying reflectivity on the transmission aperture enables communication through the frequency-doubled beam while facilitating energy transfer via the fundamental-frequency beam. In Fig.\ref{fig:doublefre}(a), a portion of the $1064\rm{nm}$ beam is initially frequency doubled to produce a $532\rm{nm}$ beam. Both beams are amplified using a gain medium, and the $532\rm{nm}$ beam is further modulated by the EOM. Subsequently, they are transmitted to the receiver with the assistance of the RIS through free space. At the receiver, a fraction of the $1064\rm{nm}$ beam passes through the output reflector and is directed to the PV panel for conversion into charging power, while the remaining portion is reflected back to the transmitter for amplification. The $532\rm{nm}$ beam, on the other hand, traverses the beam splitter and is directed to the detector for signal output.

Afterwards, the transmission power of the resonant beam and frequency-doubled beam at the transmitter, as well as the received power in the PV panel and detector at the receiver shown in Fig.~\ref{fig:doublefre}(b) can be derived by the electromagnetic wave propagation theory and the output power model. This enables the analysis of various communication performance metrics such as bit error rate (BER) and outage probability, considering different PS ratios and modulation modes in future research. More importantly, RIS deployment in free space can improve the system performance and increase the movement range for receiver.

\begin{itemize}
  \item Impact of RIS Size
\end{itemize}

\begin{figure}[!t]
\vspace{-1.0em}
\setlength{\abovecaptionskip}{0pt}
\setlength{\belowcaptionskip}{-10pt}
\centering
\subfigcapskip=-5pt
\subfigure[Different transmission distances]{
\includegraphics[width=0.31\textwidth]{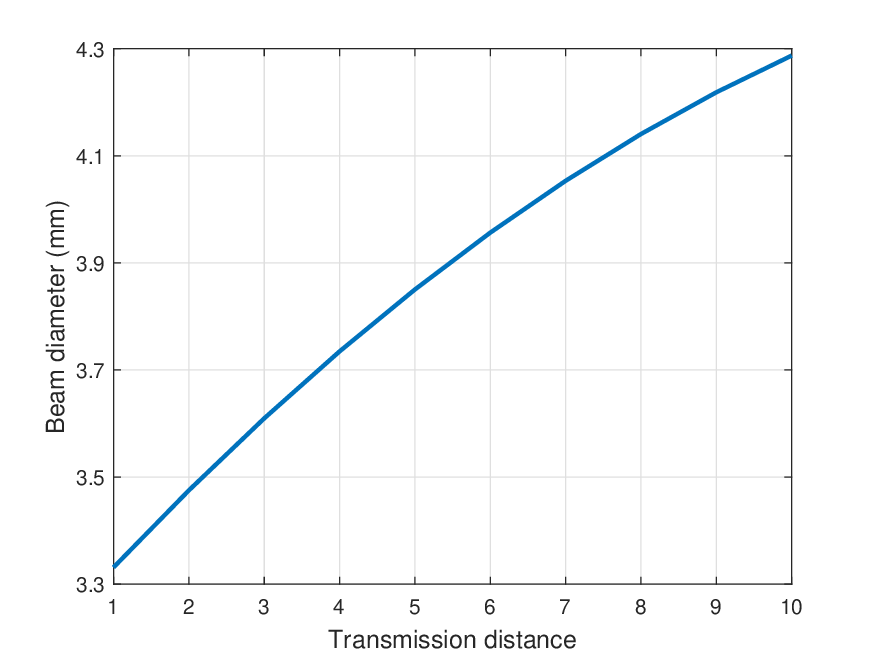}
\label{fig:beamdis}}
\subfigure[Different horizontal positions]{
\includegraphics[width=0.31\textwidth]{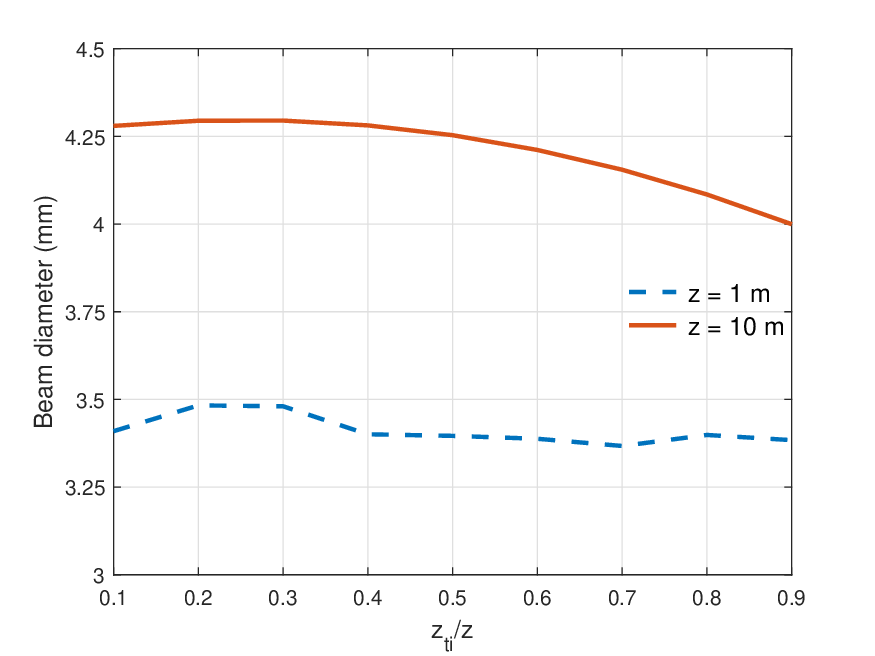}
\label{fig:beamhori}}
\subfigure[Different vertical heights]{
\includegraphics[width=0.31\textwidth]{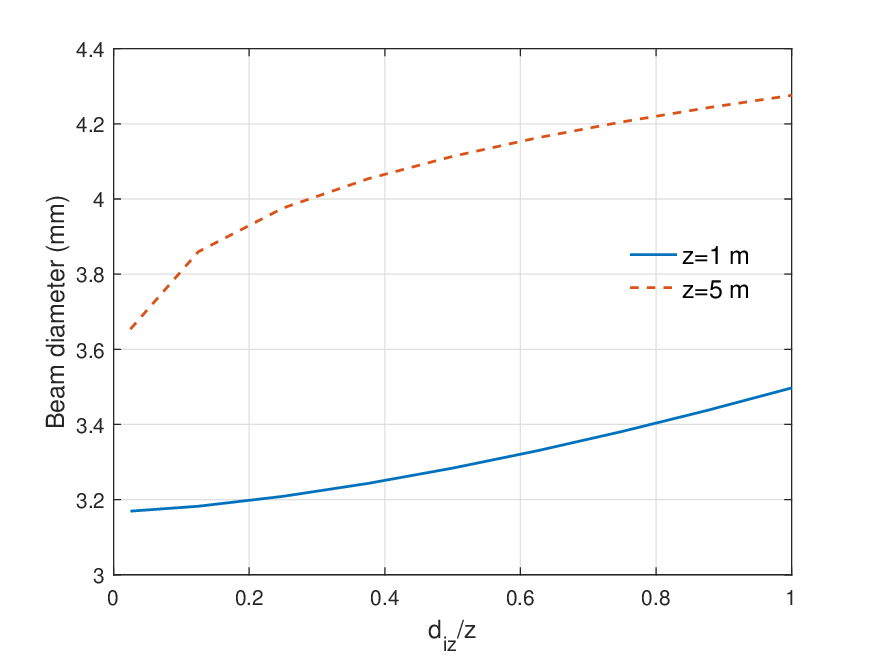}
\label{fig:beamheight}}
\caption{The changes of beam diameters on gain medium with different transfer distances and RIS placements.}
\label{Fig:beamdia}
\vspace{-2.0em}
\end{figure}

The beam spot size undergoes a process of diffusion in free space and focuses on reaching the reflectors during transmission. Meanwhile, the spot size is closely related to the transmission distance while being limited to the transfer aperture. The variations in beam diameter within the gain medium are illustrated in Fig.~\ref{Fig:beamdia}. As depicted in Fig.~\ref{fig:beamdis}, an increase in the transmission distance leads to an enlargement of the beam diameter due to proportional beam diffusion over the traveled distance. And the maximum beam diameter $4.29\rm{mm}$ at $z=10\rm{m}$, $z_{\rm{ti}}=5\rm{m}$, and $d{\rm{iz}}=0.5\rm{m}$ is less than the diameter of reflectors $5\rm{mm}$ depicted in Table.~\ref{table2}. Furthermore, in Fig.~\ref{fig:beamhori}, it can be observed that as $z_{\rm{ti}}/z$ increases, the beam diameter initially experiences growth and subsequently decreases. However, the trend is not obvious due to calculation errors. Meanwhile, Fig.~\ref{fig:beamheight} demonstrates that the beam diameter enlarges with the growth of $d{\rm{iz}}/z$.

To sum up, the transmission distance and RIS location have an impact on the diameter of the resonant beam, which is constrained to the transfer aperture. Therefore, the RIS size can be designed to match the transfer aperture for numerical analysis in the SISO resonant beam transmission system. However, the size of the RIS has an impact on the number of aligned transmitters and receivers, thereby affecting transmission efficiency, output power, channel capacity, and other performance metrics for SIMO/MIMO resonant beam transmission system, which can be studied in the future.



\begin{itemize}
  \item Performance Comparison
\end{itemize}

Recently, the most mature FSO transmission technology includes VL, laser, and resonant beam. Compared with the former two, the resonant beam-based system can realize high-power, high-capacity, long-range, safe, mobile power and information transmission over free space benefitting from the spatially separated structure. The performance comparison between resonant beam and laser WPT system has been demonstrated in \cite{fang2021safety}, where the safety of resonant beam is superior to that of the laser system with the same parameters. Additionally, FSO techniques face a fundamental limitation of NLOS links between the transmitter and the receiver. Therefore, it is crucial to investigate RIS-assisted FSO transmission, including performance analysis and comparison in the existing FSO systems.

\begin{itemize}
  \item Other Issues
\end{itemize}

Furthermore, several critical issues need to be further studied. For example, investigating integrated sensing and communication (ISAC) using RIS-aided resonant beam technology, examining the deployment of RIS with diverse structures, and optimizing the performance of RIS-assisted resonant beam systems. These research areas hold significant potential for advancing the understanding and practical implementation of RIS-assisted resonant beam technology.

\ifCLASSOPTIONcaptionsoff
  \newpage
\fi

{\small
\bibliographystyle{IEEEtran}
\begin{spacing}{1.25}
\bibliography{references}
\end{spacing}
}

%




\end{document}